\documentclass[iop]{emulateapj}
\pdfoutput=1 
\begin{document}
\newcommand{\ik}{{\it Kepler}}
\newcommand{\KIC}{{{\rm{KIC}}~9632895}}
\newcommand{\etal}{{\it et~al.~}}
\newcommand{\ie}{{\it i.e.~}}
\newcommand{\eg}{{\it e.g.~}}
\newcommand{\vs}{{\it vs.~}}
\newcommand{\kms}{\mbox{$km \ s^{-1}$}}
\newcommand{\Msun}{\mbox{M$_{\sun}$}}
\newcommand{\Mjup}{\mbox{M$_{Jup}$}}
\newcommand{\Mearth}{\mbox{M$_{\earth}$}}
\newcommand{\Rsun}{\mbox{R$_{\sun}$}}
\newcommand{\Rjup}{\mbox{R$_{Jup}$}}
\newcommand{\Rearth}{\mbox{R$_{\earth}$}}
\newcommand{\Lsun}{\mbox{L$_{\sun}$}}
\newcommand{\ltsimeq}{\raisebox{-0.6ex}{$\,\stackrel
        {\raisebox{-.2ex}{$\textstyle <$}}{\sim}\,$}}
\newcommand{\gtsimeq}{\raisebox{-0.6ex}{$\,\stackrel
        {\raisebox{-.2ex}{$\textstyle >$}}{\sim}\,$}}
\def\lesssim{\mathrel{\hbox{\rlap{\hbox{\lower4pt\hbox{$\sim$}}}\hbox{$<$}}}}
\def\gtrsim{\mathrel{\hbox{\rlap{\hbox{\lower4pt\hbox{$\sim$}}}\hbox{$>$}}}}
\def\ggg{\mathrel{\hbox{\rlap{\hbox{\lower4pt\hbox{$\sim$}}}\hbox{$>$}}}}
\def\persec{s$^{-1}$}

\title{
$\KIC$ -- The $10^{th}$ {\it Kepler} Transiting Circumbinary Planet 
}

\author{
William F. Welsh, Jerome A. Orosz, Donald R. Short}
\affil{Department of Astronomy, San Diego State University,
5500 Campanile Drive, San Diego, CA 92182-1221}
\email{wwelsh@mail.sdsu.edu, jorosz@mail.sdsu.edu}

\author{William D. Cochran, Michael Endl, Erik Brugamyer}
\affil{McDonald Observatory, The University of Texas as Austin, Austin,
TX 78712-0259}

\author{Nader Haghighipour} 
\affil{Institute for Astronomy and NASA Astrobiology Institute, University 
of Hawaii-Manoa, Honolulu, HI 96822, USA} 

\author{Lars A. Buchhave} 
\affil{Harvard-Smithsonian Center for Astrophysics, Cambridge, 
Massachusetts 02138, USA, and
Centre for Star and Planet Formation, Natural History Museum of Denmark,
University of Copenhagen, DK-1350 Copenhagen, Denmark} 

\author{Laurance R. Doyle}
\affil{SETI Institute, 189 Bernardo Avenue, Mountain View, CA 94043; and
Principia College, IMoP, One Maybeck Place, Elsah, Illinois 62028}

\author{Daniel C. Fabrycky}
\affil{Department of Astronomy and Astrophysics, University of Chicago, 
5640 S. Ellis Ave., Chicago, IL 60637}

\author{Tobias Cornelius Hinse} 
\affil{Korea Astronomy and Space Science 
Institute, 776 Daedukdae-ro, Yuseong-gu 305-348, Daejeon, 
Republic of Korea, and
Armagh Observatory, College Hill,BT61 9DG, Armagh, NI, UK} 

\author{Stephen Kane} 
\affil{Department of Physics \& Astronomy, San Francisco State 
University, 1600 Holloway Avenue, San Francisco, CA 94132, USA} 

\author{Veselin Kostov} 
\affil{Department of Physics and Astronomy, Johns Hopkins University, 
3400 North Charles Street, Baltimore, MD, 21218} 

\author{Tsevi Mazeh} 
\affil{School of Physics and Astronomy, Raymond and Beverly Sackler 
Faculty of Exact Sciences, Tel Aviv University, 69978, Tel Aviv, Israel} 

\author{Sean M. Mills}
\affil{Department of Astronomy and Astrophysics, University of Chicago, 
5640 S. Ellis Ave., Chicago, IL 60637}

\author{Tobias W. A. Mueller}
\affil{Institute for Astronomy and Astrophysics, University of Tuebingen, 
Auf der Morgenstelle 10, D-72076 Tuebingen, Germany}

\author{Billy Quarles}
\affil{NASA Ames Research Center, Moffet Field, CA 94035, USA}

\author{Samuel N. Quinn} 
\affil{Department of Physics \& Astronomy, Georgia State University, 
25 Park Place NE Suite 600, Atlanta, GA 30303;
NSF Graduate Research Fellow}

\author{Darin Ragozzine}
\affil{Florida Institute of Technology, Department of Physics and 
Space Sciences, 150 W. University Blvd., Melbourne, FL 32901} 

\author{Avi Shporer$^{1}$} 
\affil{Jet Propulsion Laboratory, California 
Institute of Technology, 4800 Oak Grove Drive, Pasadena, CA 91109, USA;
and Division of Geological and Planetary Sciences, California Institute 
of Technology, Pasadena, CA 91125, USA}
\altaffiltext{1} {Sagan Fellow} 

\author{Jason H. Steffen} 
\affil{Lindheimer Fellow, CIERA, Northwestern University, 2145 Sheridan 
Road, Evanston, IL 60208} 

\author{Lev Tal-Or} 
\affil{School of Physics and Astronomy, Raymond and Beverly Sackler 
Faculty of Exact Sciences, Tel Aviv University, 69978, Tel Aviv, Israel} 

\author{Guillermo Torres}
\affil{Harvard-Smithsonian Center for Astrophysics, 60 Garden St., 
Cambridge, MA 02138}

\author{Gur Windmiller}
\affil{Department of Astronomy, San Diego State University,
5500 Campanile Drive, San Diego, CA 92182-1221}

\author{William J. Borucki}
\affil{NASA Ames Research Center, Moffet Field, CA 94035, USA}

\altaffiltext{}
{Based on 
observations obtained with the Hobby-Eberly Telescope, 
which is a joint project of the University of Texas at Austin, 
the Pennsylvania State University, Stanford University, 
Ludwig-Maximilians-Universit\"at M\"unchen, and 
Georg-August-Universit\"at G\"ottingen.}


\begin{abstract}
We present the discovery of $\KIC$b, a 6.2 \Rearth-radius planet in 
a low-eccentricity, 240.5-day orbit about an eclipsing binary. The binary 
itself consists of a 0.93 and $0.194$ \Msun \ pair of stars with an orbital 
period of 27.3 days.
The plane of the planet's orbit is rapidly precessing, and its inclination 
only becomes sufficiently aligned with the primary star in the latter 
portion of the {\it Kepler} data. Thus three transits 
are present in the latter half of the light curve, but none of the three 
conjunctions that occurred during the first half of the light curve 
produced transits.
The precession period is $\sim$103 years, and during that cycle, transits 
are visible only $\sim$8\% of the time. This has the important implication 
that for every system like $\KIC$ that we detect, there are $\sim$12 
circumbinary systems that exist but are not currently exhibiting transits.
The planet's mass is too small to noticeably perturb the binary,
consequently its mass is not measurable with these data; but our
photodynamical model places a 1-$\sigma$ upper limit of 16 \Mearth. 
With a period 8.8 times that of the binary, the planet is well outside the 
dynamical instability zone. It does, however, lie within the habitable zone 
of the binary, and making it the third of ten {\it Kepler} circumbinary 
planets to do so.
\end{abstract}


\section{Introduction}

In its quest to find habitable worlds, the {\it Kepler} Mission
(Borucki et al.\ 2010; Koch et al.\ 2010) 
has observed over 2600 eclipsing binary star systems.
Cataloged in
Pr\v sa et al.\ (2011),
Slawson et al.\ (2011), 
and Kirk et al.\ (2014), 
the vast majority of these are new systems discovered by {\it Kepler},
and have orbital periods between 1.8 hours to 670 days. A sizeable 
fraction ($\sim$20-30\%) exhibit evidence for being triple or higher 
multiplicity stellar systems
(Conroy et al.\ 2013; Rappaport et al.\ 2013;
Gies, et al.\ 2012; Orosz et al.\ in prep).
In addition, there is a rapidly growing subset where the
third body is a planet rather than a star. 
Beyond providing challenges to planet-formation theory
(e.g. Kley \& Haghighipour 2014),
these circumbinary planets are particularly important
because the orbital configurations and 3-body gravitational 
interactions allow direct and precise measurements of the mass and radius 
of the bodies.
For example, in the Kepler-34 system the relative uncertainties in the 
stellar masses and radii are less than 0.3\% (Welsh et al.\ 2012); for 
Kepler-16 the uncertainties in the planet's mass and radius are 4.8\% and 
0.34\%, respectively (Doyle et al. 2011).

To date, nine {\it transiting} circumbinary planets have been discovered,
residing in seven systems:
Kepler-16b (Doyle et al.\ 2011),
Kepler-34b and 35 b (Welsh et al.\ 2012),
Kepler-38b (Orosz et al.\ 2012b),
Kepler-47b and c (Orosz et al.\ 2012a),
Kepler-64b (Schwamb et al.\ 2013 and simultaneously Kostov et al.\ 2013), 
Kepler-413b (Kostov et al.\ 2014), and
Kepler-47d (Orosz, in prep).
The transiting nature of these planets
unambiguously confirms the presence of the third orbiting body.
However, due to dynamical interactions, a transiting circumbinary 
planet may not always transit --- in Kepler-413 three transits 
were observed with a period of $\sim$66 days, then for 800 days no 
transits were present, then five more transits were observed. 
This behavior is due to the 2.5 degree angle between the
planet and binary orbital planes which, for Kepler-413, leads to 
precession with a period of only 11 years (Kostov et al.\ 2014).

In the following sections we present the discovery of the tenth 
{\it Kepler} transiting circumbinary planet, Kepler-nnnn (KIC~9632895).
The tight constraints placed on the relative positions, velocities, and 
sizes of the three bodies by the times, durations, and depths of the eclipses
and transits allow very precise determinations of the geometric aspects
of the system. As will be shown below, the uncertainty in the planet's 
radius is only 0.57\%, and for the secondary star's radius it is 0.65\%,
making it one of the most precisely measured low-mass stars.
Like Kepler-413, this system also exhibits times when no transits are 
present during conjunctions of the planet with the binary.
In \S2 we present the observations, in \S3 we detail the photodynamical 
modeling of the observations. We present the results in \S4 and discuss 
the characteristics of the binary and the planet, the orbital dynamics
and the long-term stability. We conclude with a section describing
$\KIC$-b's status as the third circumbinary planet in the habitable zone.


\section{Observations}

\subsection{The Kepler Light Curves}

Identification of $\KIC$ as a circumbinary planet candidate was made by 
visual inspection of a subset of the {\it Kepler} eclipsing binary star 
light curves (Slawson et al.\ 2011), in particular, those with orbital 
periods greater than $\sim$1 day that show both primary and secondary 
eclipses. Once identified as a circumbinary planet candidate, the system 
was given the KOI number 3151, though this same system had previously 
been named KOI-1451 and rejected as a false-positive; thus 
KOI-1451 is formally the correct KOI number. 
Because we identified the binary as a circumbinary planet candidate,
the target was placed on Short Cadence ($\sim$1 min integration sampling) 
starting in Quarter 13 (2012 March 29).
The {\it NExScI Exoplanet Archive} lists for KOI-1451 a {\it Kepler} 
magnitude of 13.552, temperature of $T_{\rm eff}=5618$~K, and a surface 
gravity of $\log g=4.586$, while the Mikulski Archive for Space Telescopes 
(MAST) reports $T_{\rm eff}=5425$~K and $\log g=4.803$.
Note that in their study on the rate of occurrence of circumbinary planets, 
Armstrong, et al.\ (2014) independently identified 
KIC~9632895 as a circumbinary planet candidate.

The orbital period of the binary is 27.322 days and the depth of the 
primary eclipse is $\sim$ 8\%. 
The secondary eclipse is shallow, only $\sim 0.25\%$, and flat-bottomed, 
indicating a total eclipse. Since the primary eclipse is likely 
to be total (i.e.\ annular), the depth of the eclipses tells us 
that the secondary star contributes only a small fraction to the total 
luminosity and is likely a low-mass small star.

The upper panel of Fig.~1 shows a one-year long section of the light curve, 
after normalizing each Quarter with a simple cubic polynomial.
The light curve exhibits quasi-periodic variations of $\sim$0.5\% on a 
timescale of tens of days, with the largest peak-to-peak variation being 
$\sim$1.5\%. The rms of the mildly detrended PDC-MAP light curve, after 
removing all the eclipses, is 0.22\%.
We interpret these modulations as being caused by starspots on 
the primary star.

The lower panels of Fig.~1 show a sample of eclipse events, all in Short 
Cadence, along with our best-fit model curve. The leftmost is a secondary 
eclipse, followed by a planet transit, then two primary eclipses. The 
time scale in each panel is identical and nicely demonstrates the much longer 
duration of the transit than that of the eclipses. The two primary eclipses are 
consecutive in time, yet the first shows a notable deviation in its residuals.
This is the signature of a starspot on the primary star being occulted 
by the secondary star. Such events are common, but are not readily visible 
in Long Cadence eclipse profiles.

Fig.~2 shows the phase-folded eclipse profiles, based on the 
Long Cadence {\it Kepler} observations that span 1470.5 days (from times 
-46.5 to 1424.0 in BJD--2,455,000). The preliminary model fit to the binary 
star data only (no planet included) is generally excellent, though there is 
a marked increase in the scatter of the residuals for the primary eclipse. 
These are due to the secondary star covering starspots (and possibly other 
features) on the primary star. Our ELC photodynamical model, described in 
\S3, does not include starspots, thus the minimization of residuals leads 
to both high and low scatter. If we were to use this preliminary fit, we 
would need to correct for a bias in the model -- the starspots should induce 
upward-only residuals, and as this is not the case, the model is too shallow 
compared to the true eclipse depth. It is important to note that the 
starspot-occultations lead not only to residuals in flux, but also to 
shifts in the measurements of the timing of the eclipses. We return to this
point in \S3 when describing our photodynamical modeling methodology.

The center-of-eclipse times for the 49 primary and 50 secondary eclipses
were measured in two ways:
(i) by using the best-fit model eclipse profile (described in \S 3) 
as a template and sliding it to best match the individual eclipses; 
and (ii) using the technique described in Steffen et al.\ (2011) and 
Welsh et al.\ (2012), which uses a template created from a polynomial fit
to the folded eclipse profile. The results were very comparable, so we 
adopt the former method because it should be less-sensitive to 
starspot-induced deviations in the mean eclipse profile.
The mean uncertainty is 6.8 seconds for the primary eclipses
(7.6~s for the 34 Long Cadence eclipses, 4.3~s for the 12 Short 
Cadence eclipses). The shallowness of the secondary eclipses made their 
timing measurements much more difficult: the mean uncertainty is 
117.3~s (123.5~s for the 37 Long Cadence eclipses, 
99.7~s for the 13 Short Cadence eclipses). 

A linear ephemeris was derived from the eclipse times and a time series 
of Observed minus Computed ($O-C$) values was made.
When compared with the local slope of the light curve surrounding each
primary eclipse, the $O-C$ times show a significant anti-correlation.
This anti-correlation indicates that the spin of the star is prograde 
with respect to the binary orbit (Holczer et al.~2014; see also
Sanchis-Ojeda, et al.\ 2012).
Such behavior was also seen in Kepler-47 (Orosz et al. 2012a).
Using a linear fit to the primary star $O-C$ values versus the local 
slopes, the eclipse times were then statistically corrected for the 
starspot-induced timing shifts; the uncertainties in the times were 
increased (by $\sim$10~s in quadrature) to account for noise in this 
correction. The eclipse measurements made after date BJD--2455000=1015 
used Short Cadence observations, though because of the boosting of the 
uncertainties in the starspot correction, the error bars are similar 
to those of the Long Cadence $O-C$. The secondary eclipse $O-C$ 
values were not correlated with their local slope, so no correction was made.
A common linear ephemeris was then fit to both sets of eclipse times, 
and Fig.~3 shows the resulting $O-C$ diagram.
The rms of the $O-C$ residuals is 10.1~s for the primary and 100.0~s
for the secondary, and both appear consistent with noise.
This indicates that the circumbinary object has 
no measurable gravitational effect on the binary, at least on 
a timescale of a few years. Consequently, the object is of relatively 
low mass and only an upper limit on 
its mass can be robustly determined.

The {\it Kepler} Simple Aperture Photometry ``SAP'' data were used 
throughout this paper, with the exception of measuring
the starspot modulation rms amplitude.
The light curve was detrended and normalized using the method
described in Orosz et al.\ (in prep): windows around each eclipse event
were kept and the rest of the light curve discarded. The windows
were three times the width of the eclipse. Data that were inside an 
eclipse were then masked, and for each window, a 5th order Legendre 
polynomial was fit. The eclipse was then restored and the data 
normalized by dividing by the polynomial.
Any points with a {\it Kepler} pipeline Data Quality flag 
greater than 16 (indicative of some anomaly with the observation)
were omitted prior to the detrending.
The mean Combined Differential Photometric 
Precision (CDPP) over a 3-hour baseline as reported 
on MAST is 121 $\pm$ 34 ppm.
Prior to the eclipse and transit fitting,
the light curve was heavily detrended (to minimize starspot effects)
and normalized.
The rms scatter outside of the eclipse and transit events was measured 
to be 0.0159 \% (159 ppm), consistent with the CDPP value, and
was a factor of 1.17 times larger than the mean SAP error bar. This 
difference suggests that either the SAP error bars are slightly 
underestimated or there are additional high-frequency variations that 
the detrending did not remove.

\subsection{High Resolution Spectroscopy}

$\KIC$ was observed from the McDonald Observatory with the High Resolution 
Spectrograph (Tull 1998) on the Hobby-Eberly Telescope (HET),
and with the Tull Coude Spectrograph \citep{tull1995} on the
the Harlan J. Smith 2.7~m Telescope (HJST).
A total of 11 spectra were obtained in 2013, spanning 47 nights.
The HRS spectra cover a wavelength range from 4780 to 6800 \AA \
at a resolving power of $R=30,000$ and a typical continuum 
signal-to-noise (S/N) ratio of 40:1 at 5500 \AA. The data taken with 
the Tull spectrograph span the entire optical spectrum and have a 
resolving power of $R=60,000$ with a typical continuum S/N ratio 
around 20:1 at 5500 \AA. 
We determined the RVs by cross-correlating the spectra with the RV 
standard star HD~182488.
Eleven radial velocities were extracted (note: a 0.185 \kms \
zero-point offset was found between the two spectrographs). As expected
from the light curve, the secondary star is sufficiently faint that it 
was not detected in the spectra. By injecting a synthetic spectrum 
signal into the observed spectra and attempting to recover that signal, 
an upper limit of $\sim$ 4\% was found for any non-primary star light.
This upper limit is over an order of magnitude higher than the 
secondary star's light contribution estimated via the 
photodynamical modeling described in \S3, and therefore this is not 
valuable in determining the secondary star's flux contribution.
However, it is interesting with regard to the presence of any 
contaminating third light.
The radial velocities for the primary are listed in Table~1 and the 
ELC model fit (\S 3.1) is shown in Fig.~4.

Several independent spectral analyses were carried out using the
higher signal-to-noise HET spectra, including an SPC 
(Stellar Parameter Classification) analysis 
(Buchhave et al.\ 2012), an analysis similar to the one for KOI-126 
(Carter et al.~2011), 
a MOOG-based analysis 
and by fitting a theoretical template to the spectrum 
(Tal-Or et al.\ 2013).
Each gave similar results, with the best fit
$T_{\rm eff}$ ranging from 5480 - 5620 K, and metalicity
$[m/H]$ ranging from -0.10 to +0.09.
The spectra have relatively low S/N ratio for this type of
spectral analysis and the SPC method has worked reliably 
on noisy data, hence we prefer its results:
$T_{\rm eff}=5527  \pm 50$~K,
$[m/H] = +0.09 \pm 0.08$, and 
$\log{g}=4.56 \pm 0.10$. 
This surface gravity measurement was guided, but not constrained by, 
the photodynamical modeling value of $\log{g}= 4.57$. 
The agreement in metallicity between methods was not as tight as for
$T_{\rm eff}$, thus we conservatively inflate the uncertainty in both 
metallicity and $T_{\rm eff}$ to $\pm 0.1$ and $\pm 100$ K, respectively.
In addition, the projected stellar rotation velocity,
$V_{rot} \sin{i}$, was measured to be 1.9 \kms.

\subsection{Direct Imaging}
Based on the values retrieved from MAST, the seasonal mean of the 
contamination for $\KIC$ is 7.0 $\pm$ 1\%. Such contamination will
dilute the eclipse and transit depths and could significantly bias 
the inferred radii. Therefore we undertook several direct imaging 
investigations to determine the amount of contamination.

Eighteen short exposures (10~s) of the target were obtained 
in the SDSS-r filter using the LCOGT (Brown et al.\ 2013)
FTN 2m robotic telescope in 
Haleakala, Hawaii over the span of 4 nights in 2012 March. The data
were median-combined and provided an image with a FWHM of 1.70 arcsec.
According to the KIC there should be a $\Delta$Kp = 2.8 mag fainter star 
located 4.5 arcsec south of the target (KID 9632896; g-r = 4.4 mag).
This star was not detected, although fainter stars further away were 
clearly detected. The closest star to the target in the image is about 
15 arcsec away and has r=18.3 mag according to the KIC. That star is 
also the closest star to the target detected by the UBV survey of Everett 
et al.\ (2012).

To better understand any photometric contamination due to non-target light 
captured in the {\it Kepler} aperture, we observed $\KIC$ through J, H 
and Ks filters on the WIYN High-Resolution Infrared Camera (WHIRC --
Meixner et al.\ 2010) at the Kitt Peak National Observatory on 2013 Oct 19 UT.
We employed a standard five-point dithering pattern and 30~s exposure
times in each filter. Unfortunately the conditions were not photometric 
and the seeing was $\sim$0.9 arcsec. We estimated detection limits in 
each filter following the procedure of Adams et al.~(2012).
We used the IDL {\tt aper} routine to measure the contribution from the 
target's PSF in non-overlapping, concentric annuli centered on the star.
We define a detection limit as a 5-$\sigma$ signal above the measured 
stellar PSF background in each annulus. The innermost annulus in 
each filter is defined as the measured FWHM of the stellar PSF. The
detection limits are presented in Table~2.
The expected nearby star KIC 9632896 was again not detected.
A star $\sim$ 8.5 arcsec away toward the SW was detected in the J band 
(and weakly in the H band and barely in the K band), with $\sim$0.5\% the 
brightness of $\KIC$.

Finally, an even deeper J-band observation was taken with UKIRT as part 
of the J-band survey of the {\it Kepler} field for the UKIDDS Survey.
The J-band detections were converted into expected {\it Kepler} 
magnitudes via the transformations from Howell et al.\ (2012). 
The missing KIC star was again not detected. The star 8.5 arcsec 
away was detected (expected Kepmag=19.12, based on the J-mag 17.64 
observation), along with a very faint star 7.26 arcsec due south of $\KIC$, 
with an expected Kepmag of 19.85. In summary, the star reported as 
KIC 9632896 was not detected, and no stars were detected that could 
contribute any significant contamination within the Kepler aperture.

\section{Photodynamical Modeling}

Though eclipsing, $\KIC$ is a single-lined spectroscopic binary, which 
normally does not allow a full solution of the component masses. However, 
the precise times and durations of the planetary transits place strong 
constraints on the location and relative velocity of the primary star, 
which in turn constrains the mass ratio. Additionally, the light-travel 
time effect further constrains the mass ratio.
Thus a full solution for the 
eclipsing binary is possible with a photodynamical model.

\subsection{The ELC Photodynamical Model}

An upgraded ``photodynamical'' version of the ELC code (Orosz \& Hauschildt 
2000) was developed. The upgrade allows for dynamics, instead of Keplerian 
kinematics, by integrating the Newtonian equations of motion under the 
assumption of point-masses and Newtonian gravity. A 12$^{th}$ order Gaussian 
Runga-Kutta symplectic integrator, based on the code of E.\ Hairer and M.\ 
Hairer (2003) is used. 

Transits and 
eclipses are modeled using the prescription of Mandel \& Agol (2002),
replacing the Gimenez method (Gimenez 2006) formerly used in ELC.
Quadratic limb darkening is used, following the method of Kipping (2013), 
which more naturally handles correlation and limits on the coefficients.
A total of 26 parameters are specified in the model:
the five standard Keplerian orbital parameters 
for each orbit ($P,T_{c},i,e,\omega$), 
the three masses, the three radii, the stellar temperature ratio,
two quadratic limb darkening coefficients for each star, 
the longitudinal nodal angle $\Omega$ of the planet's orbit,
and four seasonal contamination levels.
Since the Keplerian parameters evolve rapidly with time, their values
presented in Table~3 are instantaneous ``osculating'' values valid at 
the reference epoch. In particular, the time of conjunction, $T_{conj}$, 
is the time of conjunction of the body and the barycenter; for the binary 
this is approximately a primary mid-eclipse time, but for the planet 
it need not be close to an actual transit. 
Furthermore, while this 
fiducial time serves to set the position of the planet in its orbit 
{\it at the reference epoch}, it does not accurately define the planet's 
position when the model is integrated away from the reference time. To 
faithfully reproduce our solution the instantaneous positions and 
velocities given in Table~4 should be used.
Although the nodal longitude of the binary is not measurable with 
our data, it must be specified; we set the angle to be zero.
In the actual fitting, ratios and other combinations of parameters are 
often more robustly constrained by the data, e.g. mass ratios, semi-major 
axis/radius, temperature ratio, $e \cos{\omega}$, and $e \sin{\omega}$,
and these are therefore used.
Both a genetic algorithm and a Markov Chain Monte Carlo optimization was 
used to explore parameter space and find the least-squares best solution,
along with the uncertainties, defined as the interval for which the 
$\chi^{2}$ is less than the minimum $\chi^{2} + 1$ for each 
marginalized parameter.

\subsection{Photodynamical Fitting Strategy}

We initially fit the normalized and detrended  {\it Kepler} Long 
Cadence light curve plus the 11 radial velocity values.
Because the normalized flat sections of the light curve far from any 
eclipse or transit contain no information on the system parameters, they
were omitted from the fitting process. 
The uncertainties in the {\it Kepler} data were increased to account for 
additional ``noise'' that occurs when a starspot is eclipsed (see Fig.~2). 
An increase of 1.44 was chosen to give a reduced $\chi^2 \equiv 1.0$,
based on a preliminary model fit to the eclipses.
Our initial fit was acceptable, yielding  a $\chi^{2}$ of 11,114.6 for 
11,139 data points.
However, we found the results unsatisfactory for the following reason:
The best fit estimate for the mass of the planet was tightly constrained
to zero mass. When the non-negative mass constraint was lifted, the preferred 
solution fell to roughly $-90 \Mearth$, at an unnerving 3-$\sigma$ confidence.
Three scenarios were considered to explain this non-physical result,
keeping in mind that any constraint on the mass of the planet arises from
the perturbation the planet induces on the binary, which would be manifest 
most readily in the eclipse timing variations (ETVs).

The first scenario presumed that the binary apsidal motion was not
being calculated sufficienly well in the model, and the disagreement 
between the predicted and measured divergence in the $O-C$ diagram was 
leading to the non-physical mass. 
For the separation and radii of the stars, classical apsidal motion due 
to tides was found to be negligible, but the precession caused by 
general relativity (GR) was potentially signicant.
This led to the inclusion of a GR correction in the model
(Mardling \& Lin 2002; see also Ragozzine \& Wolf 2009 and 
Hilditch 2001). However, the $O-C$ of the ETVs is essentially flat, 
and the inclusion of GR increases the apsidal motion rate, thus 
having the effect of nudging the planet's mass lower. 
The amplitude of the effect is nearly negligible however. Nevertheless, 
GR is included as a necessary component for a more realistic model.

The second scenario was based on the premise that the starspots on the 
primary were affecting the times of the primary eclipses. Spurious 
correlated variations in the ETV could mimic the effect of a negative mass 
planet. This led us to re-measure the eclipse times using a different method 
and more carefully correct for the starspot-induced variations (see Mazeh, 
Holczer, Shporer 2014). The re-measurements were very similar to the original 
values, giving us more confidence in their validity. Nevertheless, the 
correlation between the ETVs and the local light curve slope indicates that 
starspots are affecting the eclipse times. The first Short Cadence primary 
eclipse shown in Fig.~1 illustrates this very well. Thus the primary 
eclipses are ``contaminated'' with noise due to starspot occultations, and 
their reliability is less than ideal. To mitigate this problem, we chose 
to fit only three primary and secondary eclipses that showed very clean 
residuals. These Short Cadence data are shown in Fig.~5. With the superb 
{\it Kepler} photometry, these six eclipses are sufficient to measure all 
the geometric binary system parameters. To capture the effect of the planet on 
the binary (the only way to measure its mass in this single-planet system), 
we include as part of the data all the eclipse times. The model is then 
penalized in a $\chi^{2}$ sense if its predicted eclipse times do not match 
the observed eclipse times. (The same 1.44 boost factor on the {\it Kepler}
SAP error bars was maintained, for both the Long and Short Cadence.)
As with the radial velocity observations, the eclipse times were given equal
weight per point as the photometric data, i.e., no regularization of
the different data sets were enforced.
With this new strategy for modeling the data, the dynamical constraints 
in the data are preserved, even though the individual eclipses are not fit. 
The key advantage of this method is that {\it it uses primary eclipse 
times that have been corrected for the effect of starspots.} 
The individual eclipse profiles cannot be corrected for the effect of 
starspots without employing a prohibitively computationally-expensive 
evolving-starspot model, requiring at least a dozen additional free 
parameters for even a rather simple two-starspot characterization (i.e,
starspot latitudes, longitudes, radii, temperatures, onset times, 
lifetimes, etc.) And so, using this strategy, our photodynamical model 
now prefers a much larger mass for the planet, and while the mass is still 
close to zero, the uncertainty is $\pm$16 $\Mearth$, a much more plausible 
result. This is the methodology and solution we adopt.

For completeness we briefly mention the third scenario we considered, which
was the inclusion of a possible second planet external to the observed one. 
Naturally, the freedom such a model allows provided a much better match to 
the observations -- formally an improvement of 100 in $\chi^{2}$ when fitting 
to all of the eclipses. Both planets had small but positive mass (uncertainty 
of a few $\Mearth$), and the period of the hypothetical second planet was 
$\sim$404 days -- interesting if true, as this is a period ratio of 1.68, 
which is a 5:3 ratio. However, we discard the second planet hypothesis 
because when using only the eclipse times and three clean eclipses as 
discussed above, the improvement is only 3.3 in $\chi^{2}$, far from being 
significant when the additional 7 free parameters are taken into account

\subsection{Fitting Assessment}

Using the Short Cadence data when possible, there are 18,472 
{\it Kepler} photometric data points, plus 11 radial velocities 
and 96 eclipse times. 
As evident in Fig.~1, 4, 5 and 6, the ELC photodynamical model provides
a good fit to the eclipse profiles, times of eclipse, radial velocities,
and transits. Formally, the model fit yields a $\chi^2$ of 18725.9, 
or a reduced $\chi^2$ of 1.0093.
In particular, the model matches the transit timing variations (TTVs) 
and the transit duration variations (TDVs). The time interval 
between the first two transits is 237.35 days, while the interval 
between the second and third transits is 235.56 days. 
The variation in transit duration is even more extreme, 
changing from 5.7 hours to 12.5 hours (see Fig.~6).
Table~5 lists the observed transit times, durations, and depths,
and the predicted values from the photodynamical model.
The large TTVs and the more than a factor of two difference in the duration
of the transit are due to the ``moving target'' effect --- the motion 
of the primary star about the barycenter. Thus the TTVs and TDVs 
follow well-defined curves as functions of the orbital phase of the 
binary and constitute an unambiguous signal of a circumbinary object, 
as no known false positive can reproduce this effect.
This ``smoking gun'' signature will remain approximately true even 
in the presence of additional planets. However, if the transiting 
planet's orbit is significantly tilted with respect to the binary's 
orbit, the precession timescale may be sufficiently short that the changing
impact parameter will alter the transit durations. In some cases, 
such as in Kepler-413, and now also in $\KIC$, the orbit precesses 
such that a large fraction of the time the planet does not 
transit the primary. We return to this topic in \S 4.3.

Finally, the parameters from the photodynamical solution were input 
into the {\tt photodynam} code of Josh Carter that was used to model 
the previous {\it Kepler} circumbinary planets 
(Carter et al.\ 2011; P{\'a}l 2012) and 
yielded good matches to the light curve. 
In addition, another completely independent photodynamical code 
developed by one of us (SMM), has been used to test and confirm the 
validity of the ELC code.

\section{Results and Discussion}

While the circumbinary configuration can in principle allow very accurate 
and precise estimates of the masses and radii of the three bodies, for
$\KIC$ the secondary star's radial velocity is not measured and there are 
only three planetary transits across the primary and none across the 
secondary. (The secondary star is so faint compared to the primary that 
transits over the secondary would not be detectable with these 
{\it Kepler} data;
one such unseen transit did occur near date BJD--2,455,000 = 548.974)
The reflected light from the planet is far too feeble to allow the
detection of any occultations. And the planet does not
noticeably perturb the binary, thus its mass can only be constrained
by limits on the eclipse timing variations, light travel time effects,
precession timescales, and stability arguments, all of which are subtle.
However, the phasing of the three transits is excellent, covering nearly 
the minimum to maximum possible duration and, importantly, there are 
three conjunctions where there is no observed transit. The {\it lack} 
of these three transits is itself an important constraint.

While ELC prefers a near zero-mass planet, the uncertainty of 16 $\Mearth$ 
means that an anomalously low mass (and density) planet is not required.
Perhaps equally significant, the model demands a planet with a mass far less 
than a Jupiter mass, ruling out any possibility of the circumbinary object 
being stellar.

\subsection{The Eclipsing Binary}

The secondary star has a mass $M_2 = 0.194 \pm 0.002 \Msun$ and 
radius $R_2 = 0.214 \pm 0.001 \Rsun$, making it one of the 
lowest-mass stars with a precise ($<$5\% uncertainty) dynamical mass 
and radius determination. It is just slightly less massive than 
Kepler-16 B (0.20255 $\Msun$; Doyle et al.\ 2011;
see also Winn et al.\ 2011 and Bender et al.\ 2012),
KOI-126 C (0.2127 $\Msun$; Carter, et al.\ 2010),
which in turn is very slightly smaller than CM~Dra B,A 
(0.2141 and 0.2310 $\Msun$; Morales et al.~2009).
The ELC model provides a temperature ratio of $T_{2}/T_{1}$=0.599,
which, combined with the spectroscopically measured T$_{\rm{eff}}$ of 
5527~K for the primary, gives a $T_{\rm{eff}}$ of 3309 K for the
secondary star. 

Dartmouth stellar evolution models (Dotter et al.~2008) were compared with 
the measured mass, radius, temperature and metallicity of the stars (we 
are assuming no alpha-element enhancement), and the results are shown in 
Fig.~7, in the form of a radius versus mass diagram. The best-matching 
isochrone was found by minimizing the distance between the observed value 
and the interpolated isochrones.``Distance'' is defined as the 
quadrature sum of the uncertainty-normalized deviations between the 
isochrone and the observed masses, radii, and temperatures, for a fixed 
metalicity.
Three isochrones are shown in Fig.~7: the first is a match to the primary 
star only (solid curve) for a [Fe/H]=0.09, which is the best estimate of the 
star's metallicity. This isochrone has a best temperature of 5429 K, and an 
age of 1.0 Gyrs.
The dashed curve is a match to both the primary and secondary stars for the 
same metallicity, and has temperatures of 5427 K and 3239 K for the primary 
and secondary stars, and an age of 2.0 Gyrs. The secondary star is very 
slightly above these isochrones, though well within 1-$\sigma$, but this 
desire to have a larger radius has the effect of ``pulling'' the isochrone 
above the primary star's position. The third isochrone, shown as the dotted 
line, is a match to the primary only, now with [Fe/H]=0.0. Its age is 1.25 
Gyrs, and its temperatures is 5501~K, a better match than the [Fe/H]=0.09 
case. However, while its mass and radius are an excellent match to the 
primary, it under-predicts the radius of the secondary. This might be a 
manifestation of the common problem of low-mass stars being too large 
and too cool compared to stellar models (see e.g.\ Torres 2013).
Additional exploration of isochrones for a spread of metalicities 
showed that ages up to 2.5 Gyrs can reasonably match the observations.
We also compared Yonsei-Yale isochrones (Yi et al.\ 2001) with the 
primary star's observed mass, radius, and temperature, and found 
essentially the same results.
To summarize, we can well-match the characteristics of the two stars with 
isochrones of age 1.0-2.5 Gyrs.

The modulation in the light curve caused by starspots 
allows the rotation period of the primary star
to be measured. Using the autocorrelation function (ACF) method of 
McQuillan et al.\ (2014), the rotation period is 20.31 $\pm$ 0.47 d.
We also visually inspected and computed the ACF for individual Quarters 
to be sure that the true period was not half or twice this value.
In comparison, the orbital period is 27.32 d, and the pseudosynchronous
rotation period (Hut 1981, 1982) for the mild eccentricity of the 
orbit $e$=0.051, is 26.90 d. Thus the star's rotation is not close 
to being synchronized with its orbit, though this is not unexpected given 
the relatively large orbital period. Because the spin should synchronize 
much sooner than the orbits should circularize (at least in a 2-body 
system), the system is likely to have been born with a low eccentricity.
The predicted $V_{rot} \sin{i}$ using the measured spin period and
stellar radius, and assuming the spin inclination is aligned with the 
binary orbit inclination, is 2.08 \kms. This is in agreement with the 
observed low projected rotation velocity of the star 1.9 $\pm$ 1.0 \kms.

Walkowics \& Basri (2013) report an age of 2.31 Gyrs for $\KIC$, 
based on their measured rotation period of 19.98~d which was
determined via a Lomb-Scargle periodogram.
Using our ACF-derived rotation period and the B-V color index
(corrected for reddening), we used five different gyrochronology 
relations and found age estimates spanning 2.11 to 2.87 Gyrs
(Barnes 2007, Mamajek \& Hillenbrand 2008, Meibom et al.\ 2009,
Barnes 2010, and Epstein \& Pinsonneault 2014).
These ages are on the high end of the nominal isochrone age range, 
but they should be regarded more as upper limits because the star
may have been influenced by tidal interactions in the binary,
which are very slowly driving the spin period to match the binary 
orbital period of 27.32 d.
Thus the gyrochronology age estimates are in fairly good agreement 
with our isochrone age estimate.

Finally, the observed $\sim$0.22\% rms fluctuations in the light curve
is slightly smaller than the solar value ($\sim$0.3\%), suggesting
an older, rather than younger, age. The Ca~II H \& K lines 
were not within the wavelength range of the HET spectra, and were too 
faint in the Tull spectra, so they could not be used as an activity or age 
indicator. The lack of any emission cores and the lack of the lithium 
6708~\AA \ absorption line rules out a very young age.
 
The ability to precisely measure the stellar masses and radii, 
especially for the low mass secondary star, and the mass ratio of 
$M_{2}/M_{1} = 0.208 \pm 0.003$, make the binary worthy of a more 
thorough investigation in its own right.

\subsection{The Planet $\KIC$ b}

While the radius of $\KIC$~b is well determined at $6.17 \pm 0.04 \
\Rearth$, we cannot reliably estimate its mass since it has no 
measurable effect on the binary during the course of our observations. 
If the planet was sufficiently massive, it would cause periodic changes in 
the binary eclipse times (ETVs) on roughly half the orbital period of the 
planet. And if the planet's orbit was not coplanar, it would cause a precession 
of the binary orbit (though measurable only if the binary orbit had nonzero 
eccentricity). This precession would be most readily detectable via a difference 
in the orbital periods of the primary and secondary stars, i.e., a divergence in 
the common-period $O-C$ diagram. Finally, a periodic change in the orbital 
inclination of the binary could lead to slight changes in the primary eclipse 
depth. As none of these are seen in the {\it Kepler} photometry, we are 
only able to estimate an upper limit on the mass, via the following ways. 
First, the photodynamical model fits place a 1-$\sigma$ value of 
$\sim$16 $\Mearth$.
Second, we can attempt to match the observed ETVs, both in their  
point-to-point fluctuations and in the long-term divergence between 
primary and secondary periods. 
However, there is little correlation between the 
observed primary eclipse timings and the expected ETV signal 
that would be 
induced by a planet of any mass. Nor is there any measurable difference in
period (mainly because of the noisy secondary star eclipse times).
These lead to a best-fit mass near zero.
The observed 1-$\sigma$ upper limit on the difference in primary and 
secondary eclipse periods is 1.7~s, which leads to an upper mass estimate 
of $\sim 30 \ \Mearth$, but with a large uncertainty of 
$\sim \pm 100 \ \Mearth$.
Finally, for completeness, a limit based on the light-travel time 
effect with a phase consistent with the orbit of the planet
gives a weak upper limit of $\sim$ 17 $\Mjup$. 

We can compare the above mass estimates/limits to the empirically expected 
mass, based on the radius of the planet. Several mass-radius relations have 
been published: Lissauer et al.\ (2011), Kane \& Gelino (2012), 
Enoch et al.\ (2012), Weiss et al.\ (2013), 
Wu \& Lithwick (2013), and Weiss \& Marcy (2014). 
Assuming that these relations are applicable to circumbinary planets, 
the mass and radius pairs that fall within the range of validity of these 
relations span $\sim$ 17--43 $\Mearth$.
These masses are larger than what the ELC model or ETV analysis prefers,
though not inconsistent with the upper limits.
Given the systematic uncertainties in the 
empirical relations, the agreement is reasonably good.
Taking 32 $\Mearth$ as a 2-$\sigma$ upper mass estimate, the mean 
density is then smaller than 0.75 $g \ cm^{3}$ with 95\% confidence. 
If the mass is truly lower than 16 $\Mearth$, then the density is 
$< 0.38 g \ cm^{3}$ 
and $\KIC$ is an unusual low-mass and low-density planet.

\subsection{Orbital Dynamics}

The photodynamical model predicts the planet's (sky-projected) inclination
oscillates with a 102.8 year period, which is also the precession period 
of the planet's orbit. For the best-fit solution, the planet's 
inclination varies by 4.575 deg. Conservation of angular momentum
requires the binary's orbital plane to also oscillate, but with a miniscule
$\sim 10^{-5}$ deg amplitude -- which is not surprising as the best-fit 
planet mass is near zero.
Fig.~8 illustrates the geometry of the binary and a spatial view of 
the rapid precession of the planet's orbit.

As a consequence of this changing inclination of the planet's orbit,
the impact parameter of the planet will vary by a large amount;
over a fairly short timescale the planet prcessed from 
non-transiting to transiting, then will become non-transiting again. 
In these {\it Kepler} data the planet transits 50\% of the time:
three non-transits followed by three transits.
However, this was very fortuitous --- 
due to the relatively large mutual inclination between the orbital 
planes, the majority of the time the planet does not transit the 
star at conjunction.
Transits only occur when the impact parameter, defined as the minimum 
projected distance of the planet's {\it limb} from the center of the primary 
star during conjunction, relative to the radius of the primary, is less than 
unity. This occurs only 8.4\% $\pm$ 0.2\% of the time.
In Fig.~9 we show explicitly the sinusoidal oscillation of the 
planet's orbital inclination, along with the mutual inclination, and
the evolution of the planet's impact parameter across the primary star. 
An interesting and important implication 
is that because of observational duration limitations, 
for every system like $\KIC$ we see, there are 
$\sim$12 similar systems that we do not see.
The importance of binary-induced planetary orbital precession in 
determining circumbinary transit probabilities 
was foreseen by the prescient Schneider (1994). 

In addition to the observed transit times, durations, and depths, 
Table~5 includes predicted near-future transit times. The near-future 
timescale is by necessity --- we predict the transits will stop being 
visible after 2015 July and not return until 2066.

\subsection{Stability and Long Term Evolution}

Following Dvorak (1986) and Holman \& Wiegert (1999), we calculated the 
orbital period where the planet is highly susceptible to a dynamical 
instability. With its orbital period 8.80 times the binary period, 
$\KIC$~b is comfortably above the critical period for circumbinary 
instability: $P_{p}/P_{crit} = 2.41$.
Unlike previous {\it Kepler} circumbinary planets, 
$\KIC$~b is not skirting on the edge of the instability.
Note: We have found an error in previously reported values for
$P/P_{crit}$ for Kepler 34, 35 (Welsh et al.\ 2012) and 38 
(Orosz et al.\ 2012b). The reported values, 1.21, 1.24, and 1.42
should be 1.49, 1.34, and 1.30.

The above analysis suggests that the orbit of $\KIC$ b is dynamically 
stable. However, it is important to note that this critical stability 
limit was derived assuming the planet has negligible mass compared 
to the binary, and is initially on a circular and {\it coplanar}
orbit. The perturbing effect of the planet on the dynamics of the binary 
is ignored. So to truly examine the long-term stability of the 
orbit of the planet, one has to integrate the equations of 
motion of the three-body system.

We first investigated the stability of the planet by applying the 
MEGNO (Mean Exponential Growth factor of Nearby Orbits) criterion 
(see Cincotta \& Simo 1999, 2000;
Go{\'z}dziewski, et al.\ 2001;
Go{\'z}dziewski \& Maciejewski 2001; and
Hinse et al.\ 2010).
MEGNO is used to determine general regions of dynamical instability,
chaotic zones, and the locations of orbital resonances. 
Fig.~10 shows the result of computing the 
MEGNO\footnote{The computation of MEGNO maps made use of the 
MECHANIC software available at
https://github.com/mslonina/Mechanic; see
S{\l}onina, Go{\'z}dziewski \& Migaszewski 2015).}
factor over a grid of the test planet's initial semi-major axis and 
eccentricity. 
All other orbital parameters are fixed and taken as the 
best-fit ELC values in Table~3. 
Each of the 240,000 pixels in the map was integrated for 
$\sim$18,500 binary periods (500,000 days). 
The color coding corresponds to the degree to which the orbit of the test 
planet is chaotic (yellow) or quasi-periodic (blue). The location of the 
planet's best-fit osculating Keplerian semi-major axis and eccentricity 
(from Table~3) is marked with the black dot.
The MEGNO map shows that the planet orbit resides in a region 
of phase space that is not chaotic for at least the $\sim$1370 yr 
duration of the MEGNO integration.
Note that the yellow instability gaps dipping into the blue region of the 
map correspond to n:1 mean-motion resonances between the planet and the
binary orbit. These resonances play an important role in establishing the 
final orbit of the planet during its migration from outer regions of 
the disk where it is presumably formed
(Kley \& Haghighipour 2014).

To better determine the long-term stability of the planet, we integrated 
the orbits of the three-body system for 100 kyrs (365 million days). 
The results are shown
in Fig.\ 11. As expected, the variations in the semi-major axis and  
eccentricity of the planet are negligibly small over the course of the 
integration, demonstrating that the orbit of the planet is stable on at 
least this timescale.
The long-term integrations also show that the mutual inclination 
between the planet and the binary is not caused by the gravitational 
interactions, i.e., it is a ``free'' inclination that is presumably 
primordial (tides are unable to significantly modify this inclination). 
Similarly, as mentioned before, the eccentricity of the binary is also likely 
primordial. The planet's eccentricity is dominated by the forcing of the 
binary, but there is a free eccentricity component as well 
(see Leung \& Lee 2012).


\subsection{A Planet in the Habitable Zone}

The habitable zone (HZ) around a binary star is neither spherical nor
fixed -- it revolves with the binary. Nevertheless, time-averaged
approximations are useful for a preliminary investigation of the orbit 
of the planet with respect to its location in the habitable zone.
In this approximation, we use the {\it time-averaged} distance of bodies 
in elliptical Keplerian orbits (see Williams 2003).
We also ignored the secondary star's flux contribution since it emits 
less than $\sim$0.3\% of the light in the {\it Kepler} bandpass. 
Using $T_{\rm{eff}}$ of 5527 K for the primary star and $T_{\rm{eff}}$ 
of 3309 K for the secondary star, the secondary emits 0.85\% of the  
bolometric luminosity. Assuming a Bond albedo of 0.34, appropriate for 
both the gas giant planets in our Solar System and for the Earth 
itself, and assuming re-emission over a full sphere, we find the 
planet's $T_{\rm{eq}}$ = 247 K and the time-averaged insolation to be 
S = 0.94 in Sun-Earth units. This is within the conservative HZ limits 
defined by the runaway and maximum greenhouse criteria for the 
inner and outer boundaries. 
Even at the most extreme ranges of star-planet separation due to the 
eccentricity, $T_{\rm{eq}}$ only varies between 237~K and 259~K.

We also calculated the properties of the HZ using the more detailed 
methodology presented by Haghighipour \& Kaltenegger (2013) and the 
Multiple Star HZ calculator\footnote{\tt http://astro.twam.info/hz/} 
(Mueller \& Haghighipour 2014). 
The results are presented in Fig.~12, where the dark green region 
corresponds to the narrow (conservative) HZ and the light green 
corresponds to the nominal (extended) HZ as defined by 
Kopparapu et al.\ (2013a,b, with coefficients updated on-line in 
2014). The dashed circle represent the dynamical stability limit.

We then computed the instantaneous insolation using our photodynamical 
model, which gave a mean insolation over one full precession cycle
of 0.958 S.
Fig.~13 shows this varying insolation incident upon the planet from 
both stars as a function of time. The most obvious variation is due 
to the orbital motion of the primary star. Superposed on this is the 
variation due to the eccentric orbit of the planet. Finally, on a 
$\sim$103 year timescale, the effects of the oscillation of the mutual 
inclination are apparent. The overall fluctuation in insolation 
is 8.2\% (rms) about the mean.
Interestingly, from the perspective of the planet, not every 
stellar conjunction is seen as an eclipse. This is due to the mutual 
inclination of the orbits. Only when the planet crosses the plane 
of the binary (i.e.\ near the nodal points) will an eclipse be seen
from the planet. Since the ratio of orbital periods is 8.8025,
and there are two nodal crossings per orbit, the eclipses are visible
every $\sim$4.4 binary periods, but varies slightly with the oscillation 
of the mutual inclination.

Although $\KIC$~b is a 6-\Rearth size planet and itself unlikely to harbor 
conditions suitable for life, such a planet could host a large moon capable 
of sustaining life. Since no evidence for such a moon is apparent we do not 
pursue this further,except to mention that in general, the effect noted by 
Mason et al.~(2013) should be considered: in a binary, tides can 
significantly change the spin evolution -- and thus the activity -- of the 
stars, rendering the system more, or less, habitable depending on the 
specifics of the binary.

Other than the outer planets of the Kepler-47 system, all transiting 
{\it Kepler} circumbinary planets fall within a factor of 2 of the critical 
radius for instability. As remarked upon in Orosz et al.\ (2012b), whether 
this is a selection effect or not, the observed close-t$O-C$ritical orbits 
have an interesting consequence: the {\it Kepler} circumbinary planets tend
to lie close to their HZ. Both Kepler-16 b and Kepler-47 c are in the HZ. 
$\KIC$~b now joins this group of HZ circumbinary planets, and currently
3 out of 10 {\it Kepler} circumbinary systems reside in the HZ.
Thus, although difficult to detect because of their large variations 
in transit timing and duration, the search for smaller Earth-size 
planets in circumbinary environments is a particularly exciting
endeavor.

\acknowledgments 

We thank Amy McQuillan of Tel Aviv University for assistance with the 
measurement of the ACF, and Justice Bruursema for assisting with the 
WIYN observations. 
W.F.W.\ thanks the Institute for Astronomy and the NASA Astrobiology 
Institute at the University of Hawaii-Manoa for their support and kind 
hospitality during his sabbatical visit when part of this project 
was carried out.
W.F.W.\ and J.A.O.\ gratefully acknowledge support from the National 
Science Foundation via grant AST-1109928, and from NASA's Kepler 
Participating Scientist Program (NNX12AD23G) and Origins of Solar 
Systems Program (NNX13AI76G).  
T.C.H.\ gratefully acknowledges financial support from the Korea Research 
Council for Fundamenta Science and Technology (KRCF) through the Young 
Research Scientist Fellowship Program and financial support from KASI 
(Korea Astronomy and Space Science Institute) grant number 
2013-9-400-0/2014-1-400-06. Numerical computations were partly carried
out using the SFI/HEA Irish Centre for High-End Computing (ICHEC) and the 
KMTNet computing cluster at the Korea Astronomy and Space Science 
Institute. Astronomical research at the Armagh Observatory is funded by 
the Northern Ireland Department of Culture, Arts and Leisure (DCAL).
N.H.\ acknowledges support from the NASA ADAP grant NNX13AF20G, NASA
Origins grant NNX12AQ62G, Astrobiology Institute under Cooperative 
Agreement NNA09DA77 at the Institute for Astronomy, University of Hawaii, 
and HST grant HST-GO-12548.06-A. Support for program HST-GO-12548.06-A 
was provided by NASA through a grant from the Space Telescope Science
Institute, which is operated by the Association of Universities
for Research in Astronomy, Incorporated, under NASA contract NAS5-26555.
T.M.\ gratefully acknowledges support of from
the European Research Council under the EU's Seventh
Framework Programme (ERC Grant Agreement No. 291352).
B.Q.\ gratefully acknowledges the support of the NASA Postdoctoral 
Program.
J.H.S.\ acknowledges the support from NASA's Kepler Participating 
Scientist Program (NNX12AD23G).
The authors acknowledge the outstanding work of David Ciardi 
(NExScI/Caltech) in organizing and maintaining the 
{\it Kepler Community Follow-up Observing Program} (CFOP) 
website\footnote{https://cfop.ipac.caltech.edu/home/}.
We also thank Phil Lucas for organizing the UKIRT J-band observations 
of the {\it Kepler} field available on the CFOP website.
This research has made use of the NASA Exoplanet Archive, which is 
operated by the California Institute of Technology, under contract with 
the National Aeronautics and Space Administration under the Exoplanet 
Exploration Program.
Kepler was competatively selected as the 10th mission of the Discovery 
Program. Funding for this mission is provided by NASA, Science Mission 
Directorate.

\clearpage
\begin{deluxetable}{crccc}
\tablecaption{Radial velocities of $\KIC$\label{RVs}}
\tablewidth{0pt}
\tablehead{
\colhead{HJD} &
\colhead{RV$_1$} &
\colhead{uncertainty} &
\colhead{Telescope} \\
\colhead{(- 2,455,000)} &
\colhead{(km s$^{-1}$)} &
\colhead{(km s$^{-1}$)} &
\colhead{}
}
\startdata
 987.95667 &  -12.065  & 0.095  & HJST Tull\\
 990.93694 &   -2.892  & 0.054  & HJST Tull\\
 991.93040 &    0.151  & 0.040  & HJST Tull\\
1000.96729 &    0.525  & 0.033  & HET HRS \\
1002.97067 &   -4.559  & 0.036  & HET HRS \\
1009.94932 &  -18.663  & 0.025  & HET HRS \\ 
1013.95060 &  -15.701  & 0.034  & HET HRS \\
1016.93457 &   -7.233  & 0.028  & HET HRS \\
1019.94404 &    1.749  & 0.028  & HET HRS \\
1022.90513 &    6.082  & 0.038  & HET HRS \\
1034.89098 &  -15.616  & 0.036  & HET HRS \\
1727.97974\tablenotemark{a}
           &   -4.774  & 0.085  & HJST Tull\\
\enddata
\tablenotetext{a}{
The bulk of our analysis was completed before this observation was made,
so it was not used. The predicted value from the photodynamical model for
this datum (which is 693~d after the last radial velocity measurement
we used) differs by only 0.093 \kms \ ($1.1 \sigma$) from the observed value.
}
\end{deluxetable}


\begin{deluxetable}{cccccc}
\tablecaption{WIYN/WHIRC Detection Limits\tablenotemark{a}}
\tablewidth{0pt}
\tablehead{
\colhead{ } &          
\colhead{ } &          
\colhead{ } &          
\colhead{$\Delta(mag)$}&          
\colhead{ } &          
\colhead{ } \\ 
\colhead{Filter}           &          
\colhead{ FWHM [\arcsec]} &          
\colhead{1.5\arcsec} &          
\colhead{2.0\arcsec} &          
\colhead{2.5\arcsec} &          
\colhead{5.0\arcsec} 
}
\startdata
J  & 0.87 & 2.6 & 3.8 & 4.1 & 4.3 \\
H  & 0.97 & 2.1 & 3.3 & 3.5 & 3.6 \\
Ks & 0.87 & 2.8 & 3.2 & 3.3 & 3.4 \\
\enddata
\tablenotetext{a}{Tabulated values give the detection limits in relative 
magnitudes at four radial distances from the source.}
\end{deluxetable}

\clearpage

\begin{deluxetable}{ccll}
\tabletypesize{\scriptsize}
\tablecaption{$\KIC$ System Parameters\tablenotemark{a} \label{param}}
\tablewidth{0pt}
\tablehead{
\colhead{Parameter}   & 
\colhead{Best fit}    &
\colhead{Uncertainty} & 
\colhead{Units}  
}
\startdata
{\underline{Binary}} &  &  & \\
$M_1$     & 0.934     & $\pm$ 0.010  & $\Msun$ \\
$M_2$     & 0.1938    & $\pm$ 0.0020 & $\Msun$ \\
$R_1$     & 0.833     & $\pm$ 0.011  & $\Rsun$ \\
$R_2$     & 0.2143    & $\pm$ 0.0014 & $\Rsun$ \\
orbital period    & 27.322037   & $\pm$ 0.000017  & days \\
$T_{\rm{conj}}$   & -34.574013  & $\pm$ 0.000060  & BJD\tablenotemark{b} \\
inclination       & 90.275      & $\pm$ 0.052     & degrees \\
$e\ \sin{\omega}$ & -0.0506     & $\pm$ 0.0037    & \\
$e\ \cos{\omega}$ & -0.006338   & $\pm$ 0.000016  & \\
eccentricity      & 0.0510      & $\pm$ 0.0037    & \\
arg. periastron   & 262.86      & $\pm$ 0.48      & degrees \\
semi-major axis   & 0.18479     & $\pm$ 0.00066   & au \\
limb darkening\tablenotemark{c} 
 \hfill           $q1_1$   & 0.413  & $\pm$ 0.055 & \\
 \hfill primary   $q2_1$   & 0.331  & $\pm$ 0.041 & \\
 \hfill secondary $q1_2$   & 0.40   & $\pm$ 0.11  & \\
 \hfill secondary $q2_2$   & 0.37   & $\pm$ 0.13  & \\
$T_1$       & 5527     & $\pm$ 100      & K\tablenotemark{d}\\
$T_2/T_1$   & 0.5987   & $\pm$ 0.0010   & K   \\
$T_2$       & 3309     & $\pm$ 100      & K   \\
$\log{g_1}$ & 4.566    & $\pm$ 0.015    & cgs \\
$\log{g_2}$ & 5.0630   & $\pm$ 0.0050   & cgs \\
contamination &  0.01  & $\pm$ 0.02     &     \\ 
                &              &                
                &              \\ 
{\underline{Planet}} &  &  & \\
$M_p$        & 0.1         & $\pm$ 16.0   & $\Mearth$ \\
$R_p$        & 6.165       & $\pm$ 0.035  & $\Rearth$ \\
orbital period   & 240.503 & $\pm$ 0.053  & days \\
$T_{\rm{conj}}$  & 68.998  & $\pm$ 0.054  & BJD\tablenotemark{b}  \\
inclination & 89.4338      & $\pm$ 0.0056 & degrees \\
$e\ \sin{\omega}$ & -0.00093  & $\pm$ 0.00020  & \\
$e\ \cos{\omega}$ & -0.03785  & $\pm$ 0.0088   & \\
eccentricity      &  0.0379   & $\pm$ 0.0088   & \\
arg. periastron   & -178.59   & $\pm$ 0.29     & degrees \\
semi-major axis   & 0.7877    & $\pm$ 0.0028   & au \\
nodal longitude $\Omega$ & 2.138 & $\pm$ 0.055 & degrees \\ 
mutual inclination & 2.297    & $\pm$ 0.030    & degrees \\
\enddata
\tablenotetext{a}{
The reported Keplerian parameters are the instantaneous
(osculating) values at the reference epoch Tref=2,454,964 BJD.}
\tablenotetext{b}{date with respect to BJD--2,455,000.}
\tablenotetext{c}{quadratic limb darkening coefficients for the primary 
and secondary stars, following Kipping (2013)}
\tablenotetext{d}{spectroscopically determined}
\end{deluxetable}

\clearpage

\begin{deluxetable}{lrrr}
\tabletypesize{\scriptsize}
\tablecaption{Barycentric Initial Dynamical Parameters in Cartesian 
Coordinates\tablenotemark{a} \label{MXYZ}}
\tablewidth{0pt}
\tablehead{
\colhead{Parameter}      &
\colhead{Primary Star}   &
\colhead{Secondary Star} &
\colhead{Planet b}    
}
\startdata
$m ~~(\Msun)$ &  $ 9.337681202794083\times10^{-1}$ 
              &  $ 1.937952112922161\times10^{-1}$  
              &  $ 1.859015330241639\times10^{-7}$ \\
$x$~~ (au)    &  $-9.748504997497649\times10^{-3}$  
              &  $ 4.697105974670451\times10^{-2}$  
              &  $ 4.127928083325089\times10^{-1}$ \\
$y$~~ (au)    &  $ 1.529515890494979\times10^{-4}$  
              &  $-7.369785770015704\times10^{-4}$ 
              &  $ 8.613416639983943\times10^{-3}$ \\
$z$~~ (au)    &  $-3.189393768140941\times10^{-2}$ 
              &  $ 1.536759849360090\times10^{-1}$  
              &  $-6.871064103438431\times10^{-1}$ \\
$v_{x}$~ (au day$^{-1}$) & $ 6.623637055950572\times10^{-3}$  
                         & $-3.191484642184871\times10^{-2}$  
                         & $ 1.765737347143242\times10^{-2}$ \\
$v_{y}$ ~(au day$^{-1}$) & $ 1.002847872768725\times10^{-5}$  
                         & $-4.832118534066960\times10^{-5}$ 
                         & $ 7.562722510169518\times10^{-4}$ \\
$v_{z}$ ~(au day$^{-1}$) & $-2.091184985444550\times10^{-3}$ 
                         & $ 1.007599740591624\times10^{-2}$  
                         & $ 9.826469371378278\times10^{-3}$ \\
\enddata
\tablenotetext{a}{Valid at the reference time epoch of 
2,454,964.00 BJD (or time = -36.00 if using BJD--2,455,000).
The (x,y) axes are in the plane of the sky, and the z direction points
towards the observer.}
\end{deluxetable}



\begin{deluxetable}{ccccccc}
\tabletypesize{\scriptsize}
\tablecaption{Transits Across the Primary Star\label{transittimes}}
\tablewidth{0pt}
\tablehead{
\colhead{Cycle}             &          
\colhead{Observed Transit}  &
\colhead{Model Transit}     &
\colhead{Observed Duration} &
\colhead{Model Duration}    &
\colhead{Observed Relative} &    
\colhead{Model Relative}    \\
\colhead{Number}            &
\colhead{Time\tablenotemark{a}} &
\colhead{Time}              &
\colhead{(hours)}           &
\colhead{(hours)}           &
\colhead{Depth}             &
\colhead{Depth}   
}
\startdata
1\tablenotemark{b}  & \nodata &   69.261 & 0 & 0 & 0 & 0 \\
2\tablenotemark{b}  & \nodata &  304.781 & 0 & 0 & 0 & 0 \\
3\tablenotemark{b}  & \nodata &  542.223 & 0 & 0 & 0 & 0 \\
4  &  781.8239 $\pm$ 0.0014     & 781.8227  &
        5.69 $\pm$ 0.17         &  5.769    &
        0.0042 $\pm$ 0.0001   &  0.0048   \\
5  & 1019.1749 $\pm$ 0.0010     & 1019.1770 &
        12.50 $\pm$ 0.12        &  12.494   &
        0.0047 $\pm$ 0.0001   &  0.0052   \\
6  & 1254.7319 $\pm$ 0.0007     & 1254.7312 &
        7.12 $\pm$ 0.09         &  6.978    &
        0.0046 $\pm$ 0.0001   &  0.0051   \\
7\tablenotemark{c}  & \nodata & 1494.1848  &
        \nodata                &  7.2365    & 
        \nodata                &  0.0055    \\
8\tablenotemark{c}  & \nodata &  1732.8717 &
        \nodata                &  8.9134    & 
        \nodata                &  0.0051    \\
9\tablenotemark{c}  & \nodata &  1967.5163 &
        \nodata                &   9.7495   &  
        \nodata                &   0.0054   \\
10\tablenotemark{c} & \nodata &  2206.4545 &
        \nodata                &   3.3775   & 
        \nodata                &   0.0030   \\
   &  &  &  &  &  & \\
89\tablenotemark{d}  & \nodata &  20973.5 & 
        \nodata & \nodata & \nodata & \nodata  \\
\enddata
\tablenotetext{a}{BJD--2,455,000.}
\tablenotetext{b}{No transit occurred because the impact parameter was 
greater than unity. The times listed are the computed times of conjunction 
from the photodynamical model.}
\tablenotetext{c}{Only transits \#4, 5, and 6 were observed.
The predicted times of transit have a precision of $<$10 minutes.}
\tablenotetext{d}{The next set of visible transits after the 2015 July 02 UT
transit is predicted to start circa 2066 November 18. The prediction
is likely good to $\sim$5 hours.}
\end{deluxetable}

\clearpage


\begin{figure}
\includegraphics[scale=0.70,angle=-90]{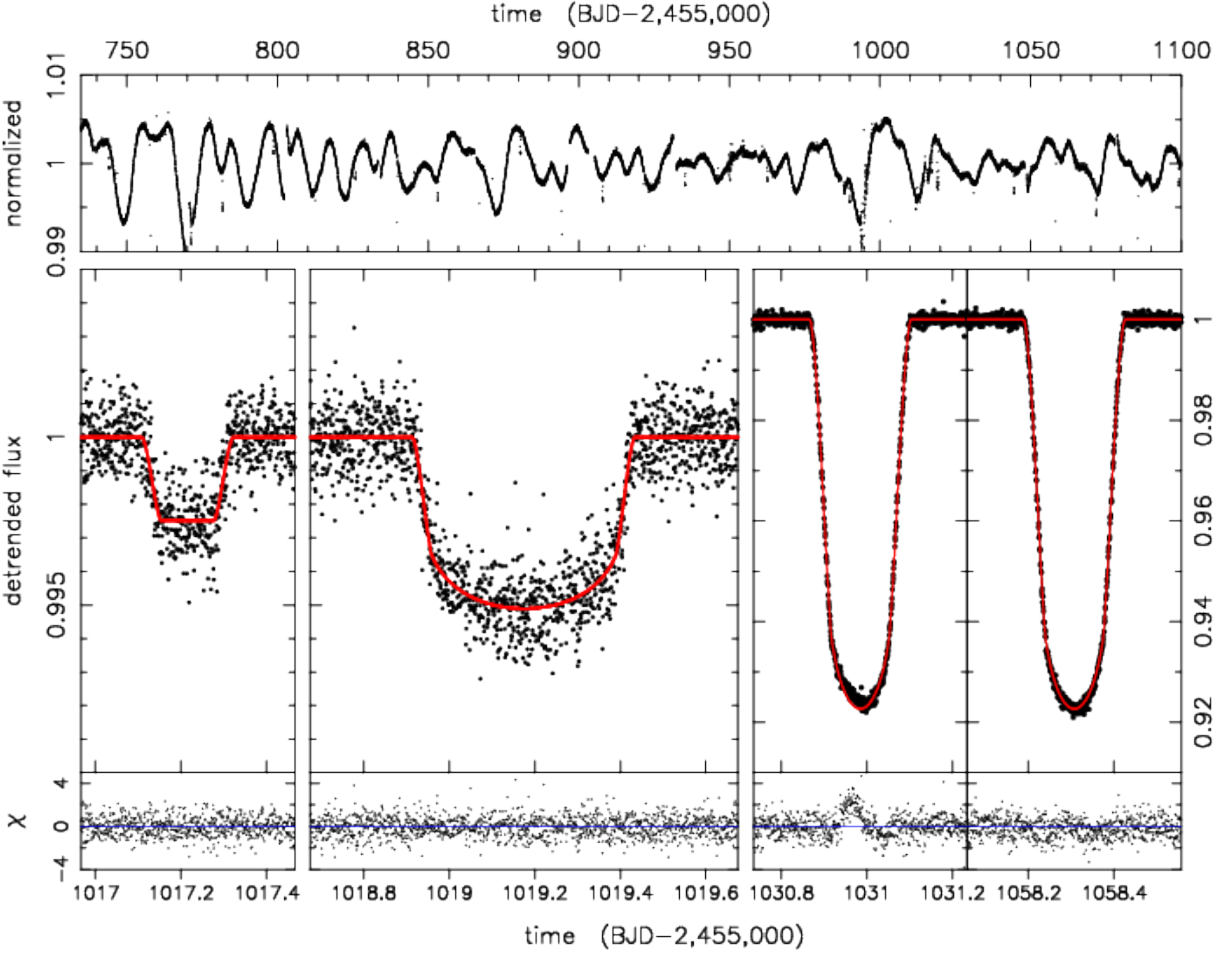}
\caption{The upper panel displays a one-year long segment of the {\it 
Kepler} $\KIC$ light curve, and shows the modulations in flux due to 
starspots. The four lower panels present examples of a secondary eclipse, 
a planet transit, and two primary eclipses, respectively, as seen in 
short-cadence (SC) data. The vertical (flux) scale is the same for the 
secondary eclipse and transit, but is different from that of the pair of 
primary eclipses. The horizontal (time) scale is the same in each panel, 
and reveals the much longer duration of the transit compared to the eclipses. 
Residuals of the ELC model fit are shown in the bottom panels, normalized 
to the uncertainty, so this is effectively a $\sqrt{\chi^{2}}$ per point. 
The first primary transit shows a clear deviation in the residuals, due to 
the secondary star crossing over a starspot.
\label{SC_fig}} 
\end{figure}
\clearpage

\begin{figure}
\includegraphics[scale=0.85,angle=0]{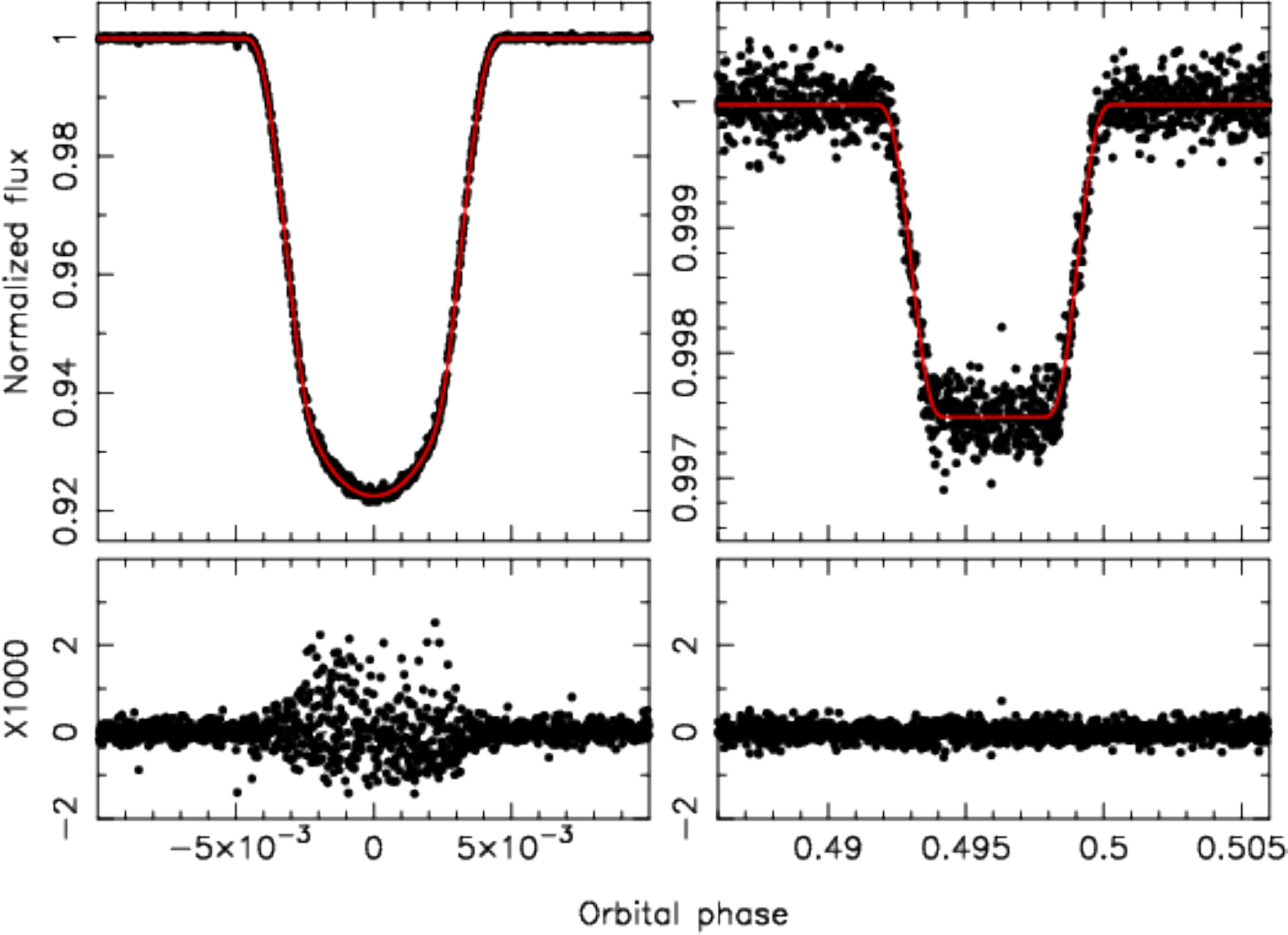}
\caption{All of the Long Cadence primary eclipses (left) and 
secondary eclipses (right) are phase-folded and plotted together. 
The offset of the center of the secondary eclipse from orbital phase 
0.5 is caused by the eccentricity of the binary.
A preliminary Keplerian model fit to the binary is overplotted on 
the data, and the residuals of the fit are shown in the bottom panel 
on a vastly enlarged scale. 
The large increase in scatter seen in the primary eclipse is due to 
spot-crossing events (i.e., the secondary covering a starspot on the 
primary).
This starspot-induced noise creates shifts in the best-fit mid-eclipse 
times and a bias in the model eclipse depth.
For these reasons we do not fit all the eclipse events, as decribed
in the text; the fit shown is not the photodynamical model used 
to determine the system parameters.
\label{foldfig}} 
\end{figure}
\clearpage

\begin{figure}
\includegraphics[scale=0.65,angle=0]{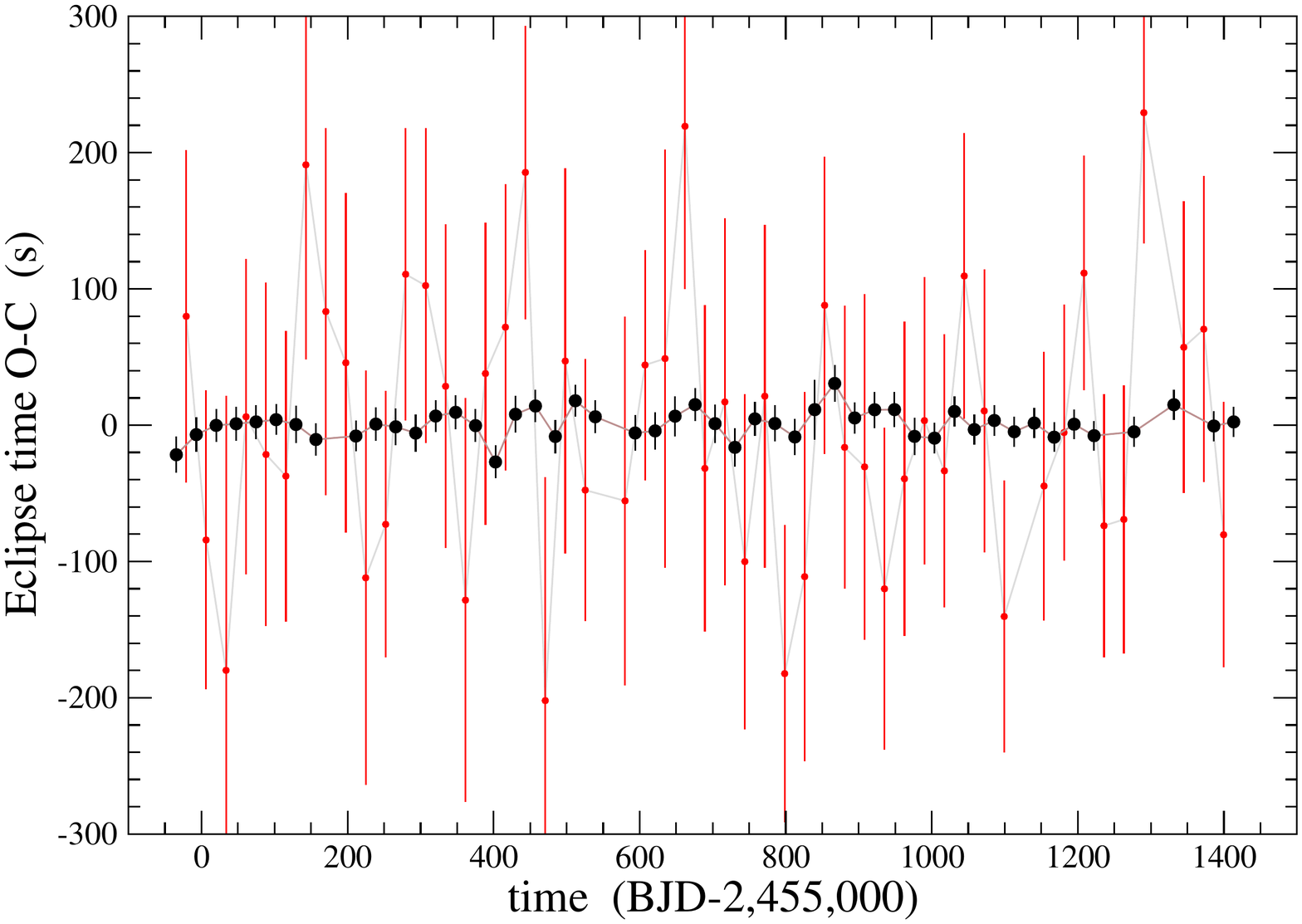}
\caption{
Starspot-corrected $O-C$ diagram showing the primary eclipse times 
(black) and secondary eclipse times (red) with respect to a common linear 
ephemeris. No trend or pattern above the noise is evident in either case.
\label{OmCfig}}
\end{figure}
\clearpage

\begin{figure}
\includegraphics[scale=0.6,angle=0]{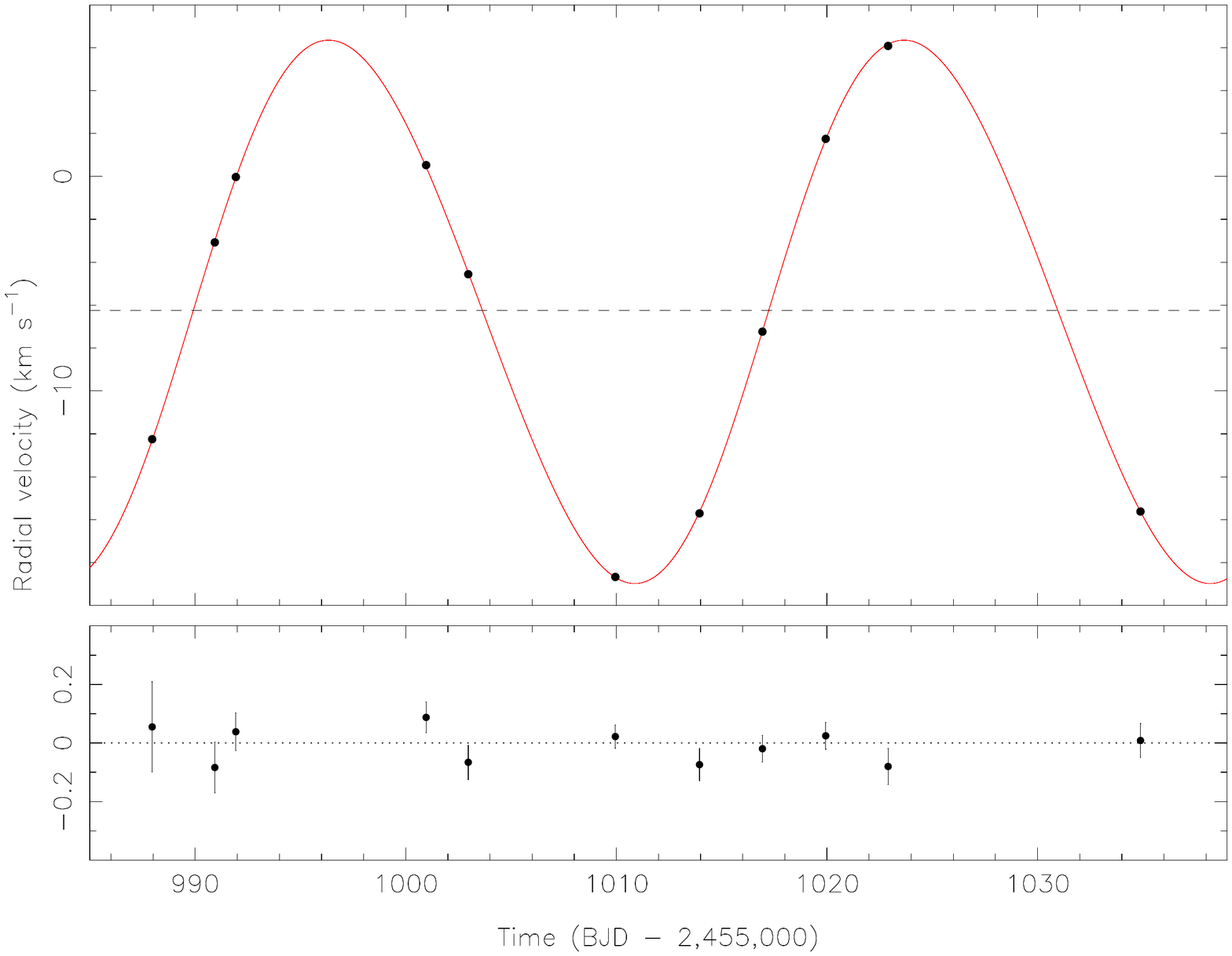}
\caption{
Radial velocities of the primary star and the best-fitting
ELC model curve, plotted versus time (not folded in phase).
Residuals are shown in the bottom panel.
\label{RVfig}}
\end{figure}
\clearpage

\begin{figure}
\includegraphics[scale=0.70,angle=-90]{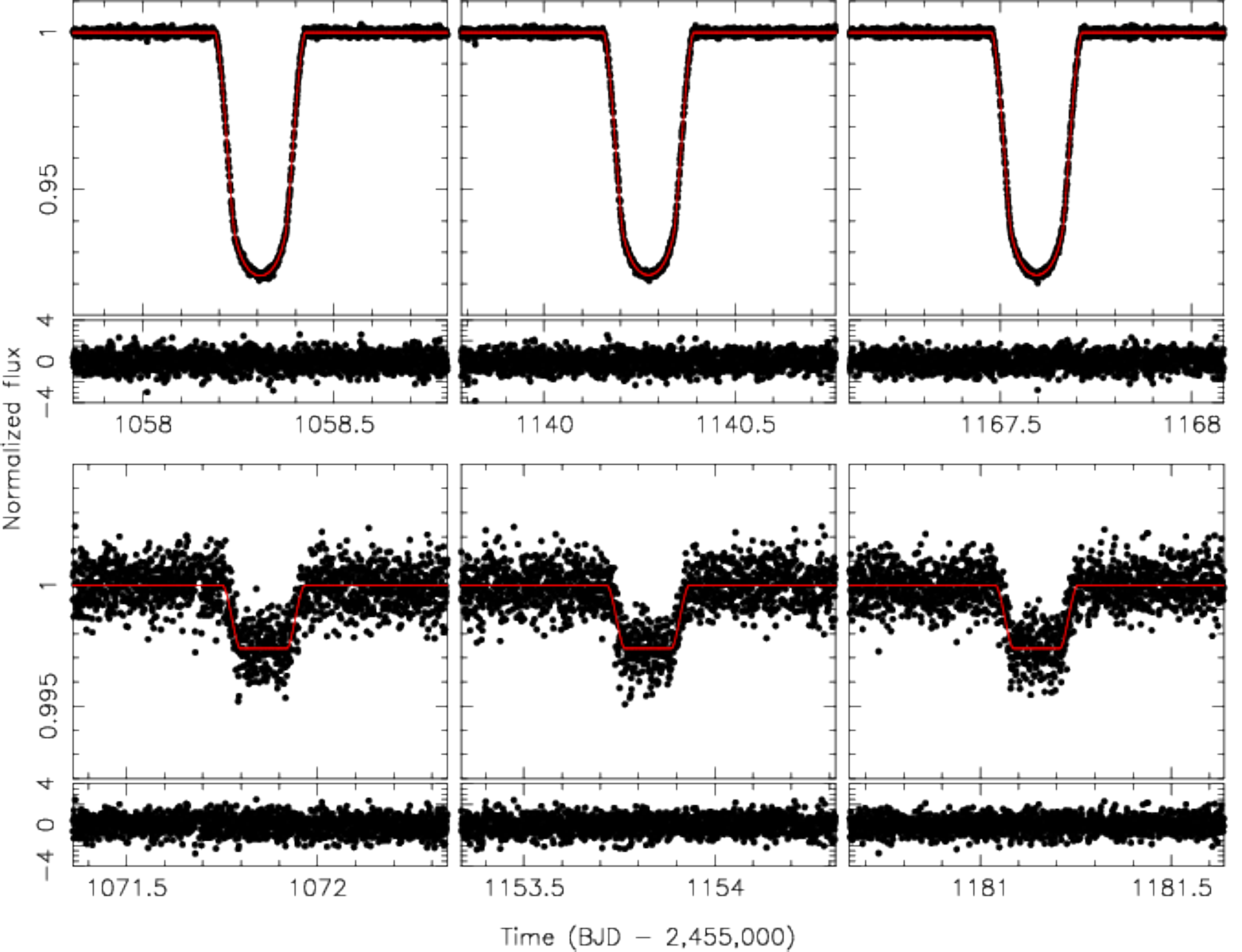}
\caption{Three primary eclipses (top row) and three secondary eclipses
(bottom row) used in the adopted fitting methodology. The best-fit
photodynamical model is overplotted in red.
Residuals of the fit are shown in the corresponding bottom 
panels, in units of parts-per-thousand.
\label{eclipsefig}}
\end{figure}
\clearpage

\begin{figure}
\includegraphics[scale=0.70,angle=-90]{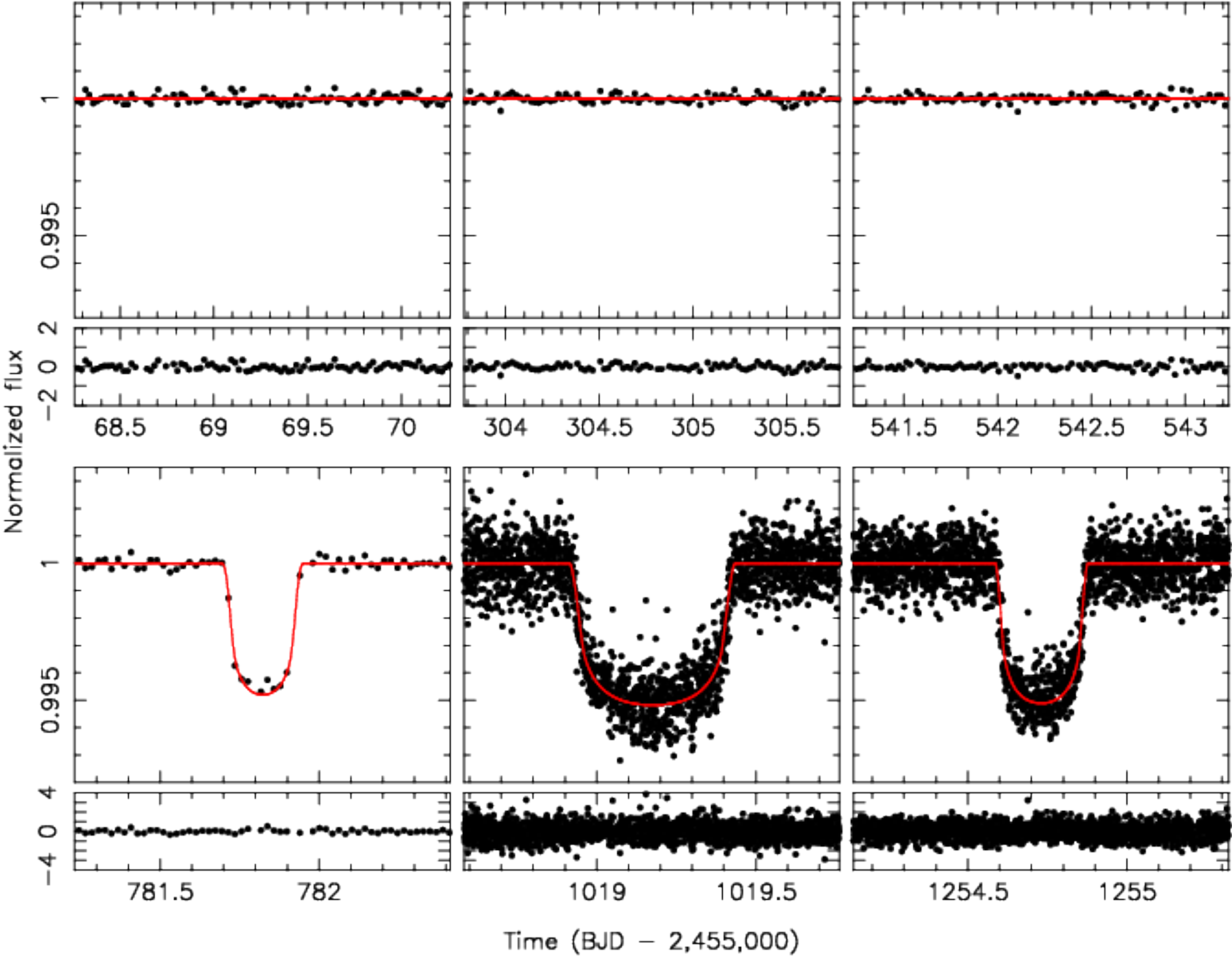}
\caption{Transit light curves and ELC model fits.
The upper panels show the epochs where transits would occur, if 
the binary and planet orbits were co-planar. The lack of transits in 
the data (and in the models) illustrate the time-varying mutual 
inclination of the orbits.
The lower panels show the three observed transits; the first transit
was observed with Long Cadence only. The large variations in transit duration 
is clear.
Note that the plot windows in the upper panels (1.5~d) are wider than 
in the lower panels (1.0~d).
\label{transitfig}} 
\end{figure}
\clearpage

\begin{figure}
\includegraphics[scale=0.70,angle=-90]{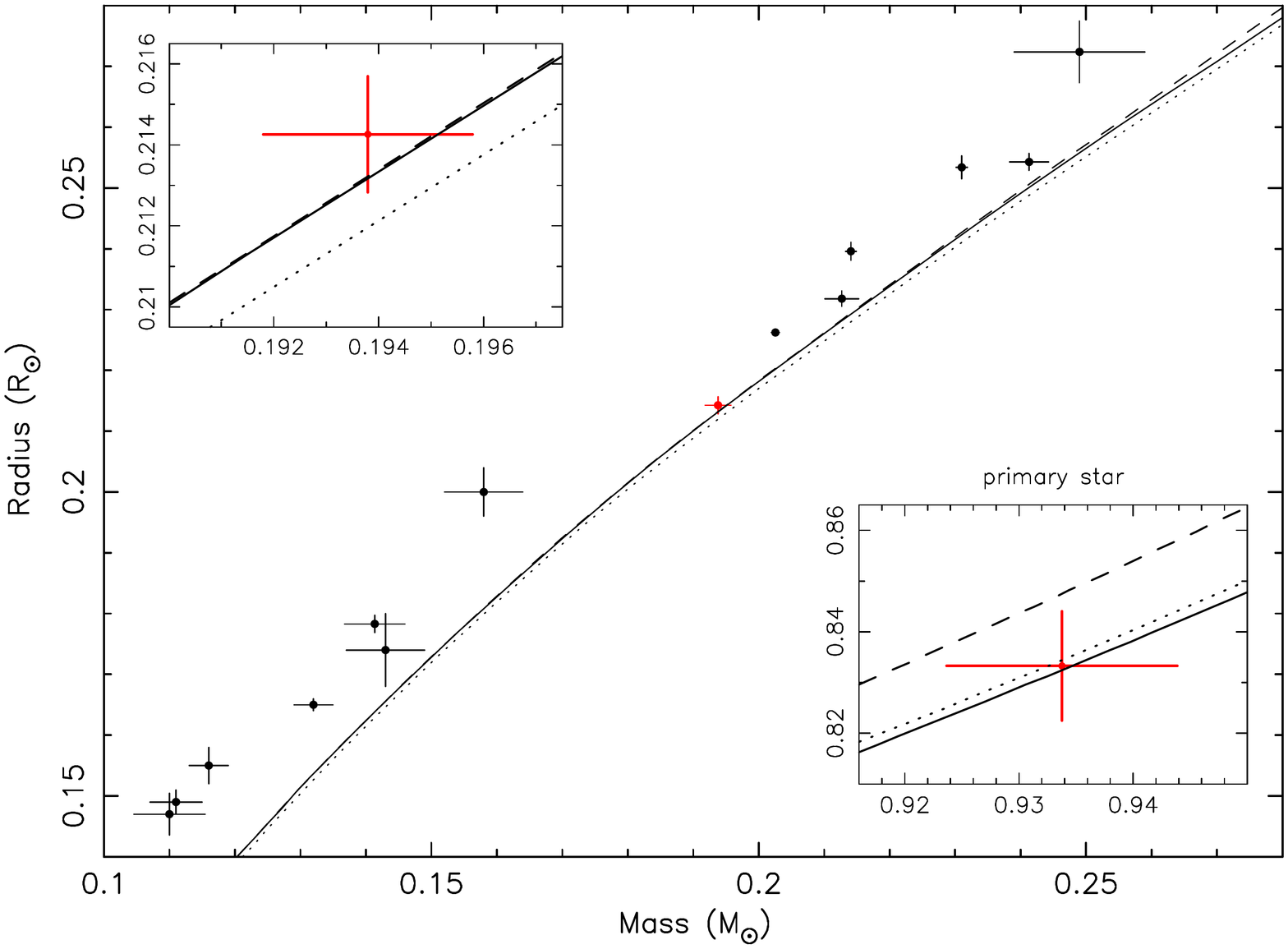}
\caption{
Mass and radius of the stars in $\KIC$, along with three Dartmouth 
isochrones (Dotter et al.\ 2008), are shown. The solid curve is the best 
match to the primary star and has an age of 1.0 Gyr; the dashed curve is 
the best match to both stars and has an age of 2.0 Gyr. Both isochrones 
have [Fe/H]=0.09, which is our best estimate for the metallicity. The 
dotted isochrone has [Fe/H]=0.00 (1-$\sigma$ lower than the measured 
value), and an age of 1.25 Gyrs. Also shown for comparison are masses and 
radii of all known low mass stars whose precision is better than 5\%.
(Note: These isochrones are not necessarily appropriate for the other
stars as their ages and metallicities are generally unknown.)
{\it Upper Inset:} A zoomed-in view for the secondary star.
{\it Lower Inset:} Similarly, but for the primary star.
\label{M-Rfig}}
\end{figure}
\clearpage

\begin{figure}
\includegraphics[scale=0.30,angle=0]{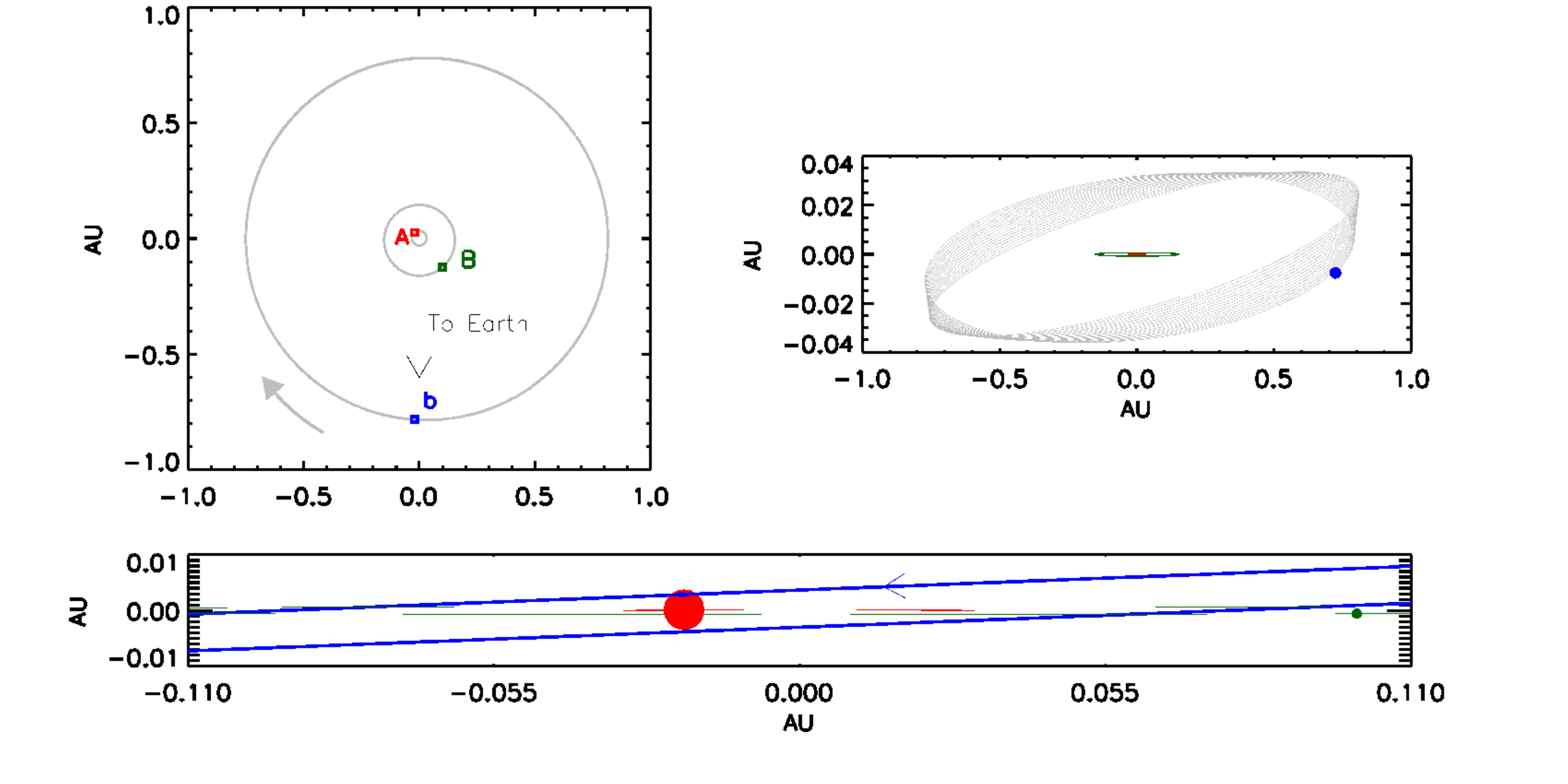}
\caption{Scaled views of the orbital configuration. The upper left 
panel shows a face-on view, and the lower panel shows the edge-on view
of the system at the first observed transit time (BJD--2,455,000 = 781.82).
The upper right panel shows the evolution of the planet's orbit from time 
6,000 to 10,000 days in BJD--2,455,000.
For clarity, the vertical scale is exaggerated by a factor of 8.
\label{orbitfig}}
\end{figure}
\clearpage

\begin{figure}
\includegraphics[scale=0.65,angle=-90]{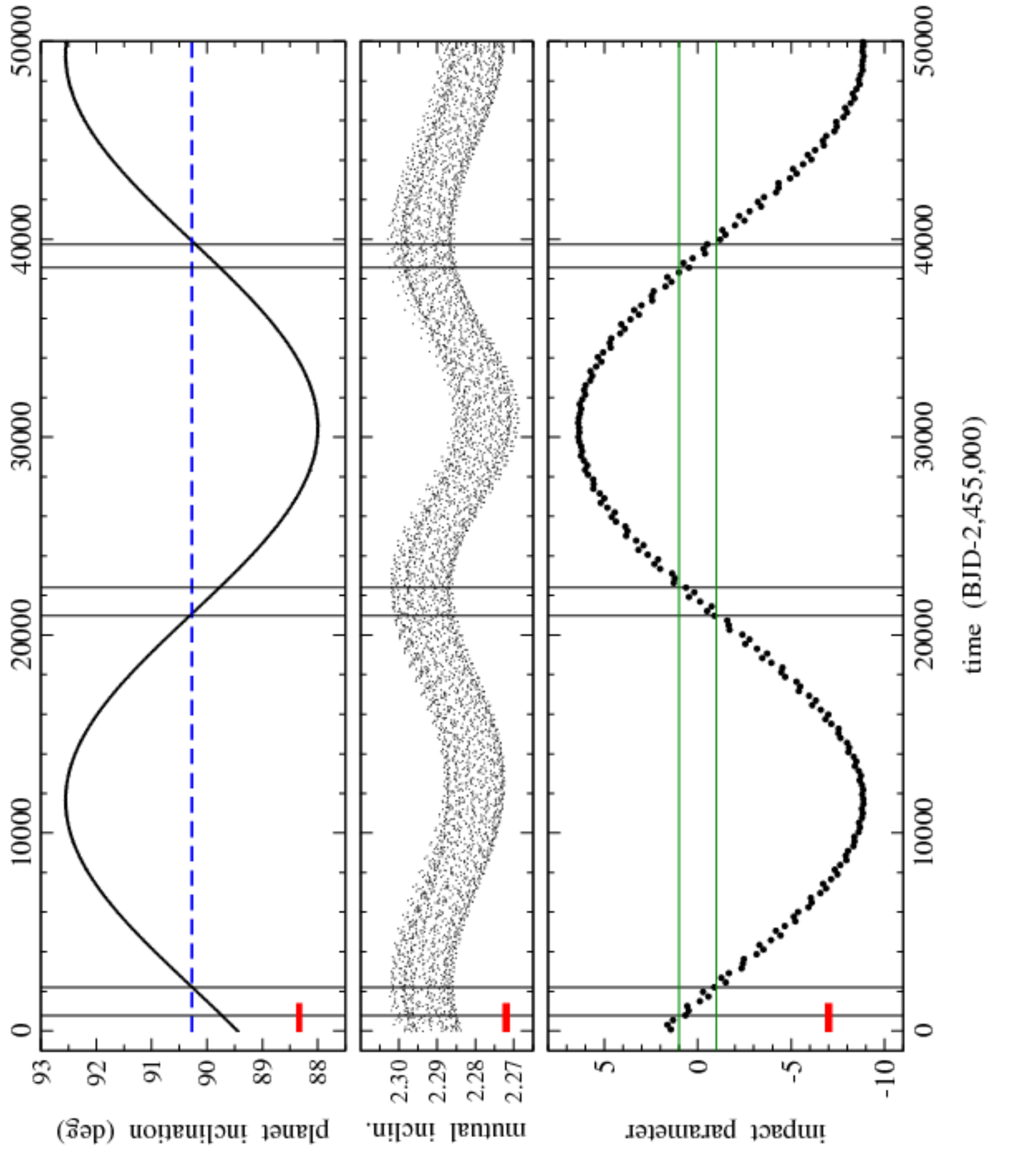}
\caption{
The upper panel shows the $\sim$103 year oscillation of the 
planet's sky-projected orbital inclination.
For comparison, the binary's orbital inclination curve is also shown 
(dashed curve) and appears completely flat on this scale.
The red horizontal marker shows the duration of the {\it Kepler} Mission.
The middle panel shows the mutual inclination of the orbits.
The bottom panel shows the variations in the impact parameter. 
Transits occur when the impact parameter is less than unity; 
this criterion is shown by the horizontal green lines. 
In each panel, the vertical lines bracket the times when the planet 
transits the primary star as viewed from Earth.
These transit windows, half a precession cycle apart, 
only encompass 8.4\% of the cycle.
\label{incfig}}
\end{figure}
\clearpage

\begin{figure}
\includegraphics[scale=0.60,angle=-90]{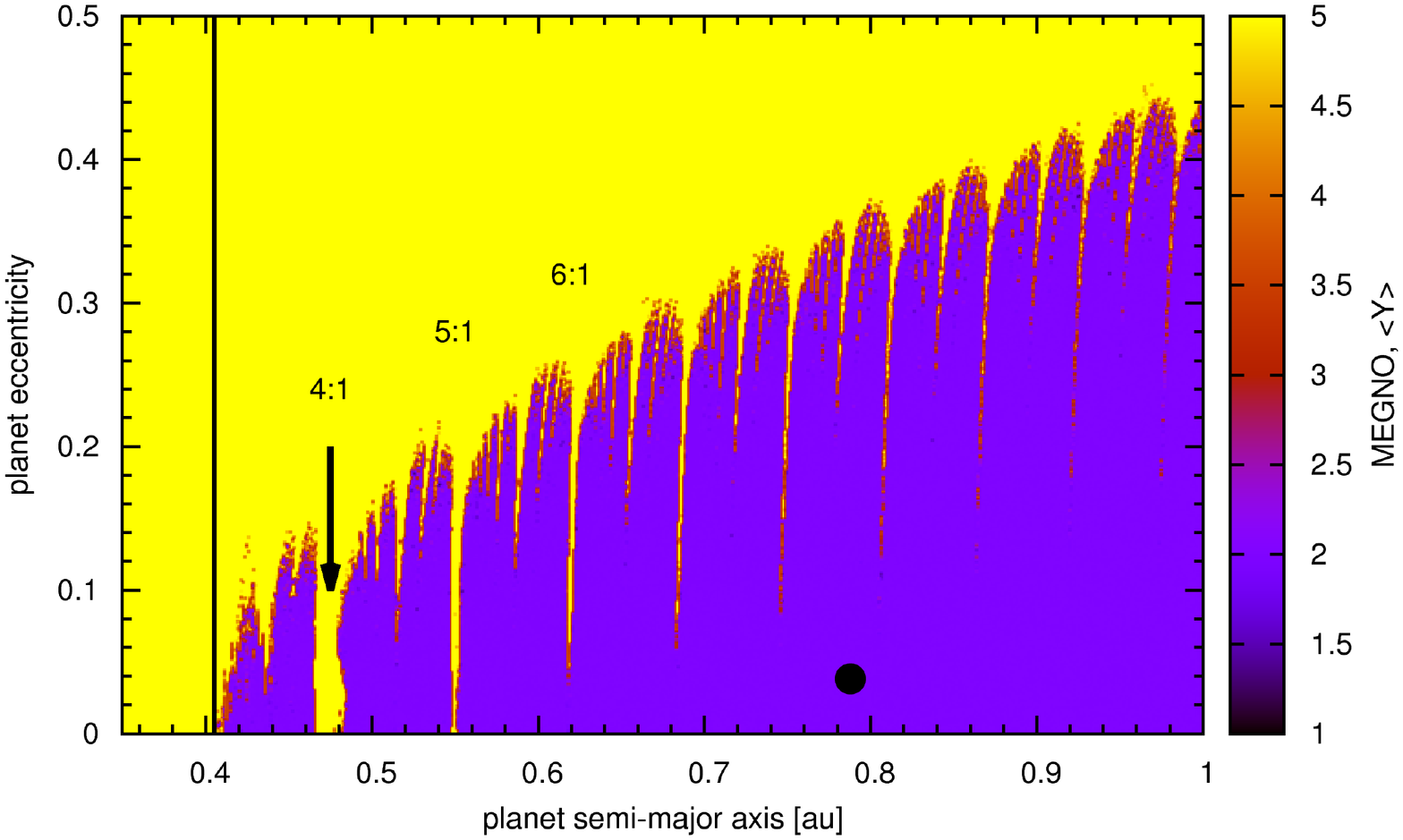}
\caption{
Map of the values of MEGNO over a grid of semi-major axis and eccentricity 
of the planet, spanning 500,000 days. The color coding indicates 
the degree of orbital instability, with yellow corresponding to chaotic 
orbits and blue depicting quasi-periodic orbits. The orbit of the 
planet resides well within a stable region.
The vertical line marks the stability limit of Holman \& Wiegert (1999)
for a zero eccentricity planet.
Note that it has been shifted from 0.437 au to 0.405 au to account for 
the difference in coordinate systems (primary centric in Holman \& 
Wiegert to Jacobi coordinates relative to the center of mass of the 
binary for MEGNO). A planet mass of 16 $\Mearth$ was used; a map with a
planet mass of zero shows no difference.
\label{MEGNOfig}}
\end{figure}
\clearpage

\begin{figure}
\includegraphics[scale=0.40,angle=0]{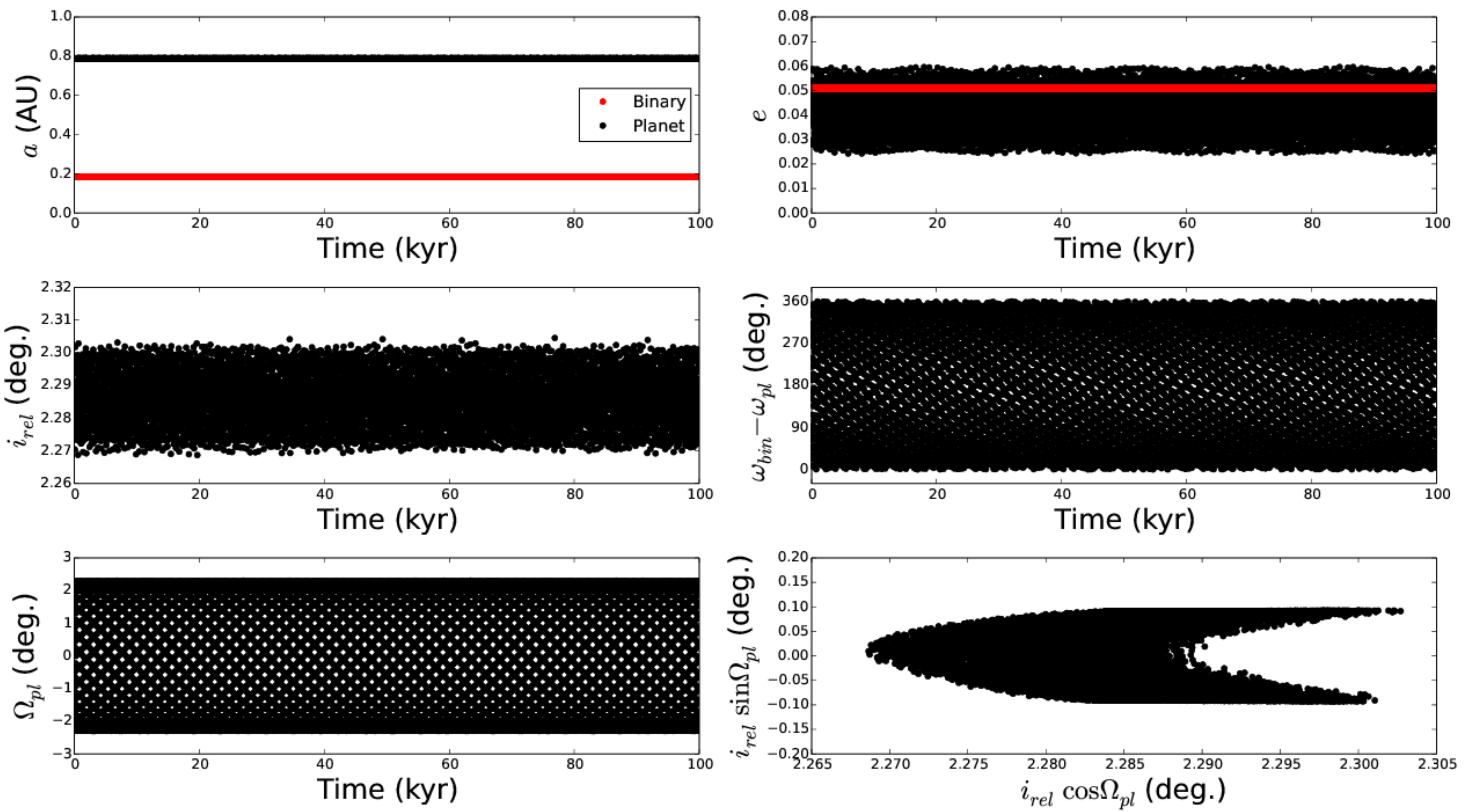}
\caption{
The 100 kyr evolution of the planet and binary orbits. 
From top to bottom the panels show the time evolution of the 
semi-major axes and eccentricities of the planet (black) and binary (red), 
the relative orbital inclination and the relative argument of pericenter,
the relative longitude of the ascending node, and the sine versus cosine
components of the $i\Omega$ vector.
For this last panel, the scale in $i \cos{\Omega}$ is ten times
smaller than in $i \sin{\Omega}$.
These figures illustrate that no secular trend is present, implying that 
the orbit of the planet is stable for long times.
\label{elementsfig}}
\end{figure}
\clearpage

\begin{figure}
\includegraphics[scale=0.7,angle=0]{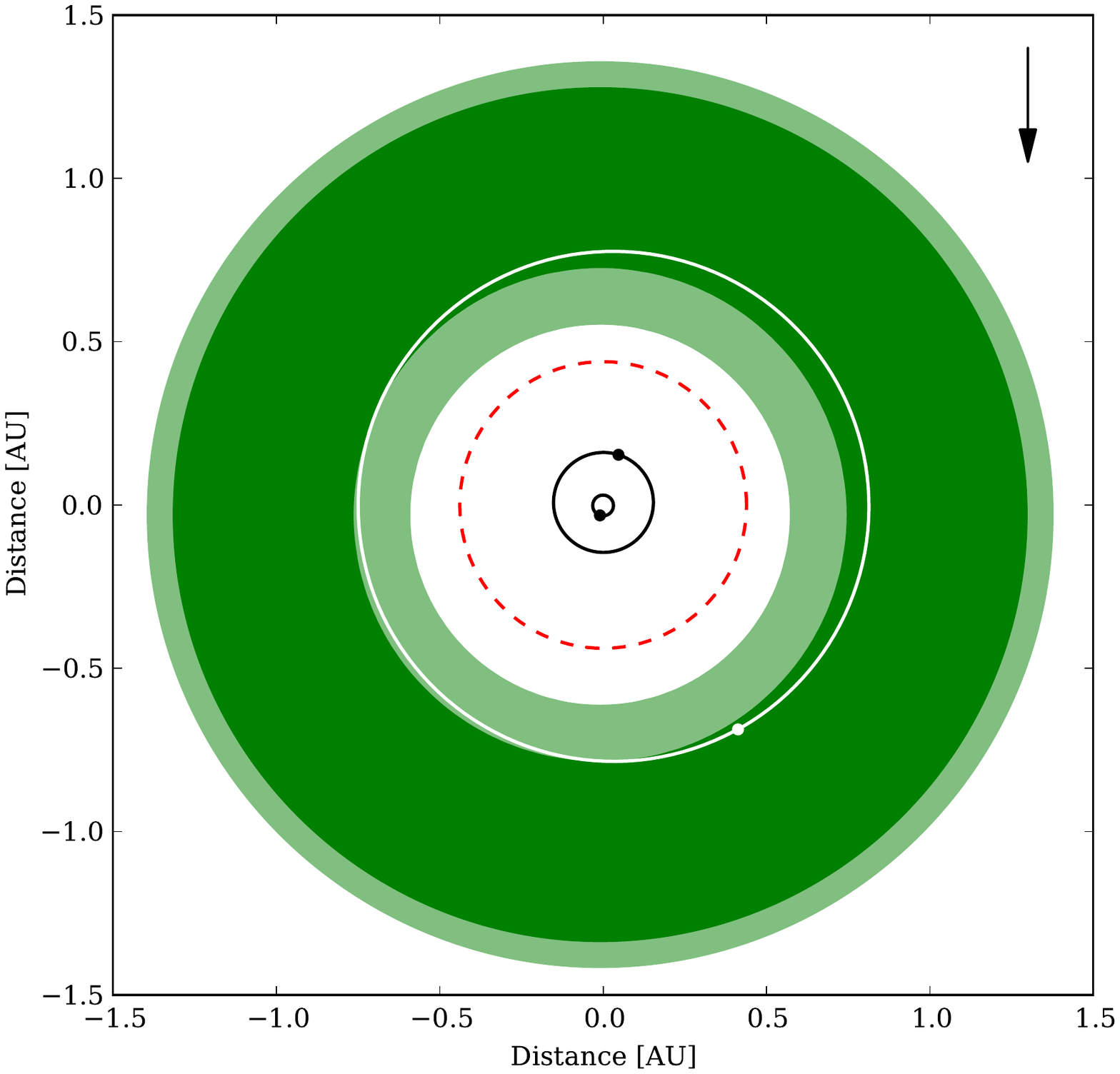}
\caption{
Face-on view of the $\KIC$ system, showing the planet's orbit relative 
to the habitable zone.
The center of the figure is at the binary center of mass, and the 
configuration corresponds to the reference epoch, with the direction of 
the line-of-sight from the Earth shown by the arrow.
The dashed red circle represents the boundary of stability for planetary 
orbits. The dark green region corresponds to the narrow (conservative) HZ and 
the light green corresponds to the nominal (extended) HZ as defined by 
Kopparapu et al (2013a,b). The orbit of the planet is shown in white. 
While the planet is likely a gas giant and not habitable, its orbit 
with respect to the HZ is of interest.
An animation of the time-variation of the HZ due to the motion of the binary 
can be found at the electronic supplementary material and at the website 
http://astro.twam.info/hz-ptype/.
\label{HZorbitfig}}
\end{figure}
\clearpage

\begin{figure}
\includegraphics[scale=0.60,angle=-90]{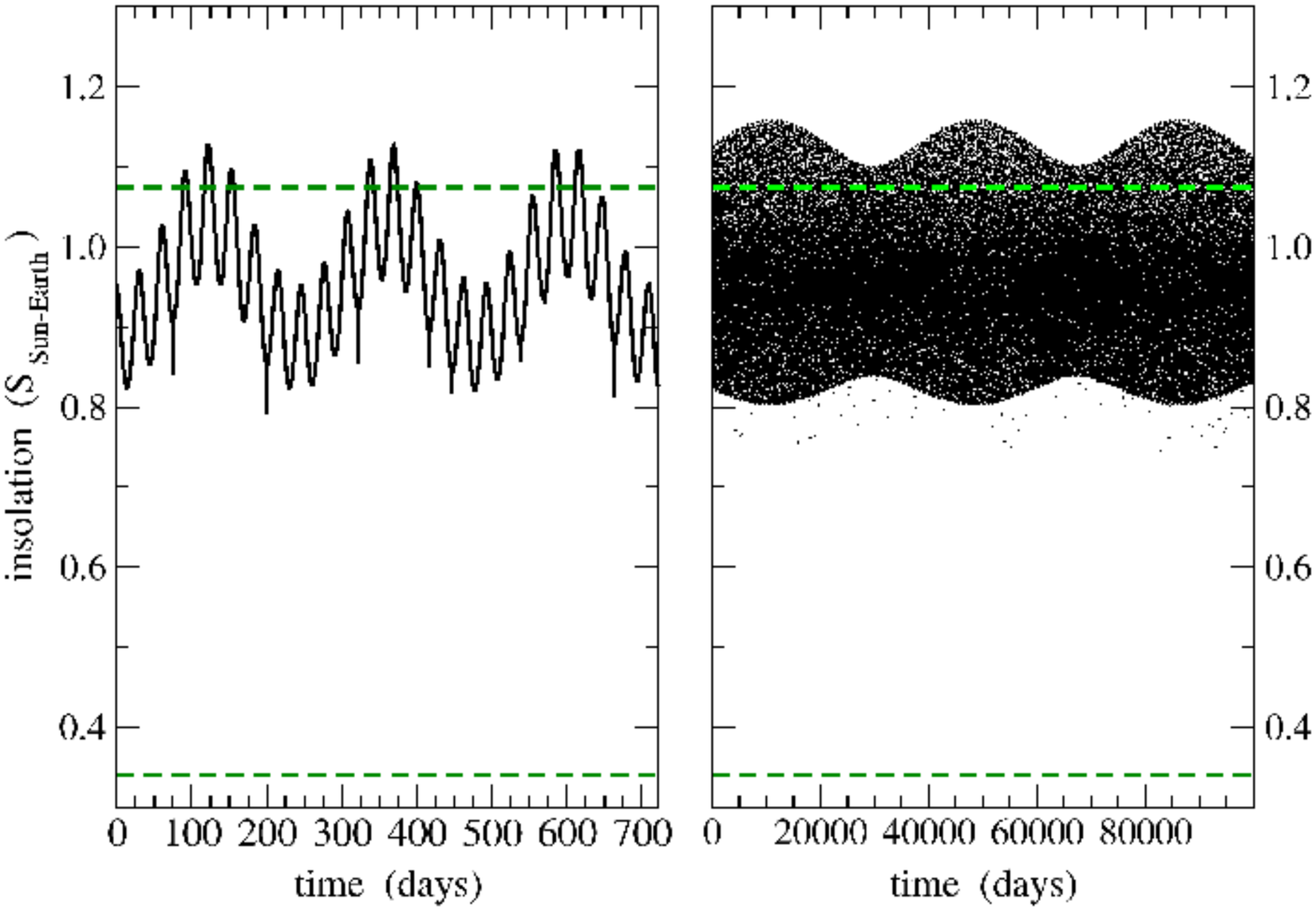}
\caption{
Sum of the fluxes (insolation) from the primary and secondary star 
incident at the top of $\KIC$~b's atmosphere.
The contribution from the secondary star is so small that it is almost 
negligible, so it is not shown.
The horizontal dashed green lines are the conservative boundaries of 
the habitable zone: the ``moist greenhouse'' (inner edge of HZ) and 
``maximum greenhouse'' (outer HZ) as defined by Kopparapu et al.\ 
(2013a,b).
{\it Left panel:} The insolation variations on short timescale, spanning
exactly three planet orbits.
The 27-d orbit of the binary is superimposed on the 240-d modulation due 
to the planet's eccentricity. The small, sharp downward spikes every 
$\sim$4th binary orbit are due to the stellar eclipses as seen from 
the planet.
{\it Right panel:} The longer term variations over a span of 100,000~d 
($\sim$274 years or $\sim$416 orbits of the planet). 
The $\sim$103-year planet precession is responsible for the sinusoidal 
envelope of the insolation.
\label{insolationfig}}
\end{figure}


\end{document}